\documentclass[preprint,12pt]{elsarticle}
\usepackage{amsmath}
\usepackage{amsfonts}
\usepackage{amssymb}
\usepackage{amsthm}
\usepackage{pgfkeys}
\usepackage{amssymb}
\usepackage{hyperref}
\usepackage{url}
\usepackage{indentfirst}
\usepackage{tikz}
\usepackage{booktabs}
\usetikzlibrary{shapes.geometric}
\usepackage{filecontents}
\usepackage{booktabs}
\usepackage[noend]{algpseudocode}
\usepackage[margin=1.5in]{geometry}    
\usepackage{algorithm}
\usepackage{graphicx}
\usetikzlibrary{shapes,positioning}
\usepackage[utf8]{inputenc}
\usepackage{hyperref}
\usepackage{caption}

\usepackage[justification=centering]{caption} 
\usepackage{graphicx}
\usepackage[export]{adjustbox}
\usepackage{caption, subcaption}

 \newcommand{\cO}{\mathcal{O}}

\newcommand{\cA}{\mathcal{A}}
\newcommand{\cB}{\mathcal{B}}

\newtheorem{theorem}{Theorem}[section]
\newtheorem{proposition}{Proposition}[section]

\newtheorem{remark}{Remark}[section]

\def\cP{\mathcal P}
\usepackage{chngcntr}
\counterwithout{remark}{section}
\counterwithout{figure}{section}
\counterwithout{table}{section}
\counterwithout{theorem}{section}
\counterwithout{lemma}{section}
\counterwithout{proposition}{section}
\journal{Discrete Applied Mathematics} 

\begin{document}
\begin{frontmatter}

\title{
Computing  
lexicographically safe Nash equilibria
in finite two-person games with tight game forms given by oracles}

\author{Vladimir Gurvich}
\ead{vgurvich@hse.ru and vladimir.gurvich@gmail.com}
\address{National Research University Higher School of Economics, 
Moscow, Russia,\\
RUTCOR, Rutgers University, Piscataway, New Jersey, United States}

\author{Mariya Naumova}
\ead{mnaumova@business.rutgers.edu}
\address{Rutgers Business School, Rutgers University, 
Piscataway, New Jersey, United States}

\begin{abstract}
In 1975 the first author proved that 
every finite tight two-person game form  $g$
is Nash-solvable, that is, 
for every payoffs  $u$ and  $w$
of two players the obtained normal form game $(g;u,w)$ 
has a Nash equilibrium (NE) in pure strategies.
Several proofs of this theorem were obtained later. 
Here we strengthen the result and give a new proof, 
which is shorter than previous ones.
We show that game  $g;u,w)$  has two types of NE, 
realized by a lexicographically safe (lexsafe) strategy  
of one player and some special best response of the other.
The prove is constructive, 
we obtain a polynomial algorithm computing these lexsafe NE.
This is trivial when game form  $g$  is given  explicitly.
Yet, in applications  $g$  is frequently realized 
by an oracle  $\cO$  such that
size of $g$ is exponential in the size  $|\cO|$  of $\cO$.
We assume that game form  $g = g(\cO)$  generated by  $\cO$  is tight 
and that an arbitrary {\em $\pm 1$  game} $(g;u^0,w^0)$
(in which payoffs  $u^0$  and  $w^0$  are 
zero-sum and take only values $\pm 1$)
can be solved in time polynomial in $|\cO|$. 
These assumptions allow us to compute 
two (one for each player) lexsafe NE in time polynomial in $|\cO|$.
These NE may coincide. 
We consider four types of oracles known in the literature and 
show that all four satisfy the above assumptions.
\end{abstract}

\begin{keyword}
Nash equilibrium \sep Nash-solvability \sep game form \sep
tightness \sep deterministic graphical game structure \sep 
game in normal and in positional form \sep
monotone bargaining \sep veto voting \sep Jordan game.

AMS subjects: 91A05, 94D10, 06E30
\end{keyword}
\end{frontmatter}

\section{Introduction} 
\label{sI} 
Here we outline main results. 
Precise definitions will be given later. 

Consider a finite $n$-person game {\em in normal form} 
representing it as a pair  $(g,u)$, 
where  $u$  is the {\em payoff} function of  $n$ players  
and  $g$ is the so-called {\em game form}. 
Respectively, the latter can be viewed as 
a game without payoffs, which are not given yet. 
Such approach is standard and convenient:  
game form  $g$ ``is responsible" for structural properties 
of game $(g,u)$, which hold for any payoff  $u$. 
For example, game form  $g$  is called {\em Nash-solvable}  
if game  $(g,u)$ has a Nash equilibrium (NE) 
in pure strategies for every payoff  $u$. 

In 1950 Nash proved that 
every $n$-person normal form game has 
a NE in {\em mixed strategies} \cite{Nas50,Nas51}. 
Yet, there are large families of games solvable  
in {\em pure strategies}, for example, 
{\em finite $n$-person positional 
(graphical) games with perfect information}. 
Its game structures $\Gamma$ uniquely defines   
a finite $n$-person game form 
$g(\Gamma$) tha is Nash-solvable.  
In Section 8.3  we consider 
this class of Nash-solvable game forms. 
Furthermore, we expand the set of outcomes 
including not only terminal positions but 
also other strongly connected components  
of the corresponding directed graph. 
Doing so, we also expand substantially 
the corresponding family of game forms, 
which remain Nash-solvable, but only in case 
of two players, $n = 2$. 

Yet, perfect information is only sufficient 
but not necessary for Nash-solvability. 
A concept  of  {\em tightness} fits much better. 
This property is of algebraic nature.   
It was introduced in \cite{Gur73,Gur75} and 
in the latter paper it was shown that  
a finite two-person game form is Nash-solvable 
if and only if it is {\em tight}. 
Note that already for  $n=3$  tightness is 
neither necessary nor sufficient for Nash-solvability. 
These results were obtained in \cite{Gur75,Gur89};  
several different proofs were given later  \cite{BBM90,BG03,DS91,Gur18,GK18}. 
Tightness remains necessary 
(and, of course, sufficient) for Nash-solvablity 
in the zero-sum case too. 
This result was obtained earlier:  
it follows easily from the so-called 
{\em Bottleneck Extrema Theorem} 
by Edmonds and Fulkerson \cite{EF70}; see also \cite{Gur73}.

Here we suggest a new 
(and much simpler) proof of the general result. 
We introduce a concept of {\em lexicographically safe (lexsafe)} pure strategy of a player 
in a given game  $(g,u)$. 
This is a refinement of the standard concept of a safe (maxmin) strategy 
that maximizes the worst possible outcome, 
while the lexsafe strategy realizes the lexicographical maximum of all possible outcomes. 
Thus, the lexsafe strategies are most safe, 
but may be not rational. 
(For comparison, recall that NE may be not Pareto optimal.)   
One can view this as a price of stability. 

We prove that a NE appears 
whenever one player applies a lexsafe strategy, 
while the opponent chooses 
some special {\em best response} to it. 
Yet, if both players choose their lexsafe strategies 
then the obtained pair may be not a NE.   

Thus, there are two types of NE: 
lexsafe for one or for the other player. 
These NE may coincide. 
For example, it happens in the zero-sum case, 
or when the considered game has a unique NE.

By definition, lexsafe strategies of a player 
do not depend on the payoff of the opponent; 
the player may be just unaware of it. 
This is an interesting property important for applications. 

In the proof given in \cite{Gur75,Gur89} 
the lexsafe strategies were  
implicitly constructed by an iterative algorithm 
increasing strategies in a lexicographical order. 
Here we suggest a simple polynomial algorithm 
searching for a lexsafe strategy of a player 
and for a corresponding NE. 
Such algorithm is obvious when 
a game form  $g$ is given explicitly. 
Yet, in applications $g$  is frequently given 
by an oracle $\cO$,   
which size may be logarithmic in the size of $g$. 
We assume that this oracle solves in polynomial time  
any two-person game  $(g,u^0)$  
in which payoff  $u^0$  is zero-sum and 
takes only values $\pm 1$; 
oracle  $\cO$  tells us who wins and 
determines a winning strategy. 
Based on this assumption, we provide an algorithm 
computing a lexsafe NE in an arbitrary game  $(g,u)$ 
in time polynomial in the size of $\cO$. 

In the last section we consider four examples of such oracles 
from different areas of game theory and 
show that all four satisfy the above assumption.

\section{Basic definitions} 
\label{s0}
\subsection{Game forms}
\label{ss01}
In this paper we consider finite,
not necessarily  zero-sum, normal form games of two players, Alice and Bob.
They have finite sets of strategies  $X$  and  $Y$, respectively.
A {\em game form} is a mapping  $g : X \times Y \rightarrow  \Omega$, where
$\Omega$  is a finite set of outcomes.
Several examples are given in Figure \ref{f1}, where
game forms are represented by tables with rows, 
columns, and entries labelled by $x \in X$, $y \in Y$, and  $\omega \in \Omega$, respectively.

\begin{figure*}[h]
\captionsetup[subtable]{position = below}
\captionsetup[table]{position=top}
\hspace*{4em}
\begin{subtable}{0.3\linewidth}
\centering
\begin{tabular}{|c|c|}
\hline
$\omega_1$ & $\omega_1$ \\ \hline
$\omega_2$ & $\omega_3$ \\ \hline
\end{tabular}
\caption*{$g_1$}               \label{tab:g1}
\end{subtable}
\hspace*{-3em}
\begin{subtable}{0.3\linewidth}
\centering
\begin{tabular}{|c|c|c|c|}
\hline
$\omega_1$ & $\omega_1$ & $\omega_2$ & $\omega_2$ \\ \hline
$\omega_3$ & $\omega_4$ & $\omega_3$ & $\omega_4$ \\
\hline
\end{tabular}
\caption*{$g_2$}
\label{tab:g2}
\end{subtable}
\hspace*{-2.5em}
\begin{subtable}{0.3\linewidth}
\centering
\begin{tabular}{|c|c|c|}
\hline
$\omega_1$ & $\omega_1$ & $\omega_3$\\ \hline
$\omega_1$ & $\omega_2$ & $\omega_2$\\ \hline
$\omega_3$ & $\omega_2$ & $\omega_3$ \\ \hline
\end{tabular}
\caption*{$g_3$}               \label{tab:g3}
\end{subtable}

\vspace{0.5cm}

\hspace*{4em}
\begin{subtable}{0.3\linewidth}
\centering
\begin{tabular}{|c|c|c|}
\hline
$\omega_1$ & $\omega_1$ & $\omega_3$\\ \hline
$\omega_1$ & $\omega_1$ & $\omega_2$\\ \hline
$\omega_4$ & $\omega_2$ & $\omega_2$ \\ \hline
\end{tabular}
\caption*{$g_4$}
\label{tab:g4}
\end{subtable}
\quad
\hspace*{-4em}
\begin{subtable}{0.3\linewidth}
\centering
\begin{tabular}{|c|c|c|c|}
\hline
$\omega_1$ & $\omega_2$ & $\omega_1$ & $\omega_2$\\ \hline
$\omega_3$ & $\omega_4$ & $\omega_4$ & $\omega_3$\\ \hline
$\omega_1$ & $\omega_4$ & $\omega_1$ & $\omega_5$\\ \hline
$\omega_3$ & $\omega_2$ & $\omega_6$ & $\omega_2$ \\ \hline
\end{tabular}
\caption*{$g_5$}
\label{tab:g5}
\end{subtable}
\quad
\hspace*{-3.5em}
\begin{subtable}{0.3\linewidth}
\centering
\begin{tabular}{|c|c|}
\hline
$\omega_1$ & $\omega_1$ \\ \hline
$\omega_1$ & $\omega_2$ \\ \hline
\end{tabular}
\caption*{$g_6$}
\label{tab:g6}
\end{subtable}

\vspace{0.5cm}

\hspace*{4em}
\begin{subtable}{0.3\linewidth}
\centering
\begin{tabular}{|c|c|}
\hline
$\omega_1$ & $\omega_2$ \\ \hline
$\omega_2$ & $\omega_1$ \\ \hline
\end{tabular}
\caption*{$g_7$}
\label{tab:g7}
\end{subtable}
\quad
\hspace*{-4em}
\begin{subtable}{0.3\linewidth}
\centering
\begin{tabular}{|c|c|c|}
\hline
$\omega_1$ & $\omega_1$ & $\omega_2$\\ \hline
$\omega_3$ & $\omega_4$ & $\omega_3$\\ \hline

\end{tabular}
\caption*{$g_8$}
\label{tab:g8}
\end{subtable}
\quad
\hspace*{-3.5em}
\begin{subtable}{0.3\linewidth}
\centering
\begin{tabular}{|c|c|c|}
\hline
$\omega_1$ & $\omega_1$ & $\omega_2$\\ \hline
$\omega_4$ & $\omega_5$ & $\omega_2$\\ \hline
$\omega_4$ & $\omega_3$ & $\omega_3$ \\ \hline
\end{tabular}
\caption*{$g_9$}
\label{tab:g9}
\end{subtable}

\caption{Nine game forms. Alice and Bob choose rows and columns, respectively.\\
Forms $g_1$ - $g_6$ are tight, forms $g_7$ - $g_9$ are not; 
see section \ref{ss04} for the definition.}
\label{f1}
\end{figure*}

Mapping  $g$  is assumed to be surjective, but not necessarily injective,
that is, an outcome $\omega \in \Omega$ 
may occupy an arbitrary array in the table of  $g$.

A pair of strategies $(x,y)$  is called a {\em situation} 
(term "strategy profile" is also used in literature).
Sets  $g(x) = \{g(x,y) \mid y \in Y\}$ and  $g(y) = \{g(x,y) \mid x \in X\}$
are called the {\em  supports} of strategies  $x \in X$  and   $y \in Y$, respectively.

A strategy is called {\em minimal} 
if its support is not a proper superset of the support of any other strategy. 
For example, in  $g_6$  the first strategies of Alice and Bob are minimal, 
while the second are not;
in the remaining eight game forms all strategies are minimal.
Moreover, any two strategies of a player have distinct supports, 
for every game form, except  $g_7$.

A situation $(x, y)$ is called {\em simple}  if
$g(x)  \cap  g(y) =  \{g(x,y)\}$.
For example, all situations of  game forms  $g_1, g_2, g_8, g_9$  are  simple
(such game forms are called  {\em rectangular});
in contrast, no situation is simple in  $g_7$;
in  $g_3$  all are simple, except three on the main diagonal;
in  $g_4$  all are simple, except the central one;
in  $g_6$  all are simple, except one with the outcome  $\omega_2$.

\subsection{Payoffs and games in normal form}
\label{ss02}
{\em Payoffs} of Alice and Bob are defined by real valued mappings
$u : \Omega \rightarrow \mathbb{R}$  and   $w : \Omega \rightarrow \mathbb{R}$, respectively.
We assume that both players are maximizers.
Triplet  $(g;u,w)$  defines a  {\em finite two-person game in normal form},
or just a {\em game}, for short.
Game $(g;u,w)$  and payoffs  $(u,w)$   are called:
\begin{itemize}
\item {\em zero-sum} if  $u + w = 0$, that is, $u(\omega) + w(\omega) = 0$  
for all  $\omega \in \Omega$;
\item {\em zero-sum $\pm 1$} 
(or just $\pm 1$, for short) if, in addition, 
functions  $u$ and  $w$  take only two values  $1$  and $-1$.
\end{itemize}

Alternatively, a $\pm 1$ payoff can be given by a partition 
$\Omega = \Omega_A \cup \Omega_B$, where 
$\Omega_A$ and $\Omega_B$  are the outcomes preferred by Alice  and by Bob, respectively:
$$u(\omega) = 1, w(\omega) = -1 \text{ for } \omega \in \Omega_A \; \text{ and } \;
 u(\omega) = -1, w(\omega) = 1  \text{ for } \omega \in \Omega_B.$$
For a $\pm 1$ game notation $(g;\Omega_A,\Omega_B)$  will be used along with $(g;u,w)$.

\subsection{Nash equilibria and saddle points} 
\label{ss03}
Given a game  $(g;u,w)$, a situation
$(x,y)$  of its  game form  $g : X \times  Y \rightarrow A$
is called a {\em Nash equilibrium} (NE)  if
$$u(g(x,y)) \geq  u(g(x',y)), \forall x' \in X,  \; \text{ and } \;
  w(g(x,y)) \geq  w(g(x,y')), \forall  y' \in Y;$$
\noindent
that is, if neither Alice nor Bob can profit
replacing her/his strategy 
provided the opponent keeps his/her one unchanged, 
or in other words, 
if  $x$  is a best response for  $y$ 
and  $y$  is a best response for  $x$. 
Note that a best response may be not unique. 

\medskip 

This concept of solution was introduced
in 1950 by John Nash \cite{Nas50,Nas51}. 
In the zero-sum case, a NE is called {\em a saddle point}. 
The latter concept was known much earlier; 
the former one is its natural extension to the non-zero-sum case.

\smallskip

Recall that a zero-sum game  $(g,u,w)$ has a saddle point 
if and only if  maxmin and minmax  are equal:    
\begin{equation}
\label{eqMMMM}
\max\min = \max_{x \in X} \min_{\omega \in g(x)} u(\omega); \;
\min\max = \min_{y \in Y} \max_{\omega \in g(y)} u(\omega);  
\end{equation}
furthermore, maxmin $<$ minmax  if and only if 
$(g;u,w)$ has no saddle point.

\begin{remark}
In \cite{Nas50,Nas51} solvability in mixed strategies is studies.
In contrast, we restrict the players to their pure strategies. 
Such approach is considered, for example, in 
\cite{AFPT10,AHMS08,BBM90,BCGM20,BFT17,BG03,BGMP10,BGMS07,FHKOY20,Gur75,Gur89,Gur17,Gur18,GK18,GN22,HKM21,Kuk11,MS96,Ros73,SM13,THS12,Was90}. 
\end{remark}

\subsection{Solvability of game forms}
\label{ss04} 
A game form  $g$  is  called: 
$(i)$ {\em Nash-}, $(ii)$ {\em zero-sum-}, $(iii)$ {\em  $\pm 1$-solvable}  
if the corresponding game  $(g;u,w)$  has a NE for 
$(i)$ all, $(ii)$ all zero-sum, 
$(iii)$ all zero-sum $\pm 1$ payoffs, respectively.

\smallskip

Implications $(i) \Rightarrow (ii) \Rightarrow (iii)$ are obvious.
In fact, all three properties are equivalent \cite{Gur75,Gur89,Gur09}.
For $(ii)$ and $(iii)$  it  was shown earlier by Edmonds and Fulkerson \cite{EF70}; 
see also \cite{Gur73}. 
The list of equivalent properties $(i)$, $(ii)$, $(iii)$
was extended in  \cite{Gur75} as follows.

\subsection{Replacing payoffs by preferences and eliminating ties}
\label{ss21}
Given a game $(g;u,w)$, we can assume wlog  that payoffs 
$u : \Omega \rightarrow \mathbb{R}$  and 
$w : \Omega \rightarrow \mathbb{R}$  have no ties. 
Indeed, one can get rid  of  all ties 
by arbitrarily small perturbations of values of $u$ and $w$.
In accordance with definition, the set of NE will be either 
unchanged or reduced by such perturbations.
We focus on Nash-solvability 
(in pure strategies), 
that is we study conditions that guarantee 
the existence of a NE for arbitrary payoffs  $u$ and $w$. 
Hence, we can wlog assume that both, $u$ and $w$, have no ties  
and replace them by linear orders 
$\succ_A$  and  $\succ_B$  over the set of outcomes $\Omega$,  
which are called the {\em preferences} of Alice and Bob, respectively. 
Thus, game  $(g;u,w)$  can be replaced by  $(g;\succ_A,\succ_B)$ 
and it is enough to study Nash-solvability  of the latter. 
Althgough some NE of  $(g;u,w)$  may disappear in $(g;\succ_A,\succ_B)$, 
yet, Nash-solvability holds or fails for both games simultaneously.

\begin{remark}
Above arguments would fail in the case of mixed strategies. 
\end{remark} 

\subsection{Tight game forms}
\label{ss05}
Mappings  $\phi : X \rightarrow  Y$  and  $\psi : Y \rightarrow X$
are called {\em response strategies} of Bob and Alice, respectively.
The motivation for this name is clear:
a player chooses his/her strategy
as a function of a known strategy of the opponent.
Standardly, $gr(\phi)$ and $gr(\psi)$ denote the graphs
of mappings  $\phi$  and  $\psi$  in  $X \times Y$.

Game form  $g : X \times Y \rightarrow \Omega$  is called {\em tight} if

\medskip

(l)  $g(gr(\phi)) \cap g(gr(\psi)) \neq \emptyset$
for any mappings  $\phi$  and  $\psi$.

\medskip

It is not difficult to verify that in Figure 1
the first six game forms ($g_1 - g_6$) are tight, 
while the last three ($g_7 - g_9$) are not.

\medskip

In \cite{EF70,Gur73,Gur75,Gur89,Gur18}
the reader can find several equivalent properties characterizing tightness.
Here we recall some of them.

\medskip

(ll-A) For every response strategy $\phi : X \rightarrow  Y$
there exists a strategy  $y \in Y$  such that  $g(y) \subseteq g(gr(\phi))$.

\medskip

(ll-B) For every response strategy $\psi : Y \rightarrow  X$
there exists a strategy  $x \in X$  such that  $g(x) \subseteq g(gr(\phi))$.

\medskip

It is not difficult to see that (l) and (ll-A) are equivalent \cite{Gur73,Gur89}. 
Then, by transposing  $g$, we conclude that (l) and (ll-B) are equivalent as well. 
Hence, all three properties are equivalent. 
One can verify this for nine examples $g_1-g_9$. 

Properties (ll-A) and (ll-B) show that playing 
a zero-sum game   $(g;u,w)$
with a tight game form  $g$ 
Bob and Alice do not need non-trivial response strategies but
can restrict themselves by the standard 
ones, that is, by  $Y$ and $X$, respectively.

\medskip

Given a  game form  $g : X \times Y \rightarrow \Omega$,
introduce on the ground set $\Omega$ of the outcomes 
two multi-hypergraphs  $\cA = \cA(g)$  and   $\cB = \cB(g)$
whose edges are the supports of strategies of Alice and Bob, respectively: 
$$\cA(g) = \{g(x) \mid x \in X\} \text{ and }  \cB(g) = \{g(y) \mid y \in Y\}.$$

Recall that distinct edges of a multi-hypergraph 
may contain one another or even coincide.
Obviously, the edges of  $\cA$  and  $\cB$  pairwise intersect, that is,
$g(x) \cap g(y) \neq \emptyset$  for all  $x \in X$  and  $y \in Y$.
Furthermore,  $g$  is tight if and only if

\medskip

(lll) hypergraphs  $\cA(g)$  and  $\cB(g)$  are {\em dual}, 
that is, satisfy also the following two (equivalent) properties:

\begin{enumerate}[{(lll-A)}]
\item for every  $\Omega_A \subseteq  \Omega$  such that
$\Omega_A \cap  g(y) \neq \emptyset$  for all  $y \in Y$
there exists an  $x \in X$  such that  $g(x) \subseteq \Omega_A$;

\item  for every  $\Omega_B \subseteq  \Omega$  such that
$\Omega_B \cap  g(x) \neq \emptyset$  for all  $x \in X$
there exists an  $y \in Y$  such that  $g(y) \subseteq \Omega_B$.
\end{enumerate}

\begin{remark}
Verification of tightness of an explicitly given game form 
is an important open problem. 
No polynomial algorithm is known. 
A quasi-polynomial one was suggested in \cite{FK96}; see also \cite{GK99}.
\end{remark} 

\section{Tightness and solvability}
\label{ss06}
Let us recall an old theorem. 

\begin{theorem} (\cite{Gur75,Gur89}) 
\label{t-NS}
The following properties of a game form are equivalent: 
$(i)$ Nash-, $(ii)$ zero-sum- , $(iii)$ $\pm 1$-solvability, 
and (iv) tightness.
\end{theorem}

\proof 
As we already mentioned, implications 
$(i) \Rightarrow (ii) \Rightarrow (iii)$ are obvious.

\smallskip

Also $(iii) \Rightarrow (iv)$ is easily seen. 
Indeed, assume for contradiction that a game form $g$ is not tight. 
Then, there exists a response strategies  
$\phi : X \rightarrow Y$  and  $\psi : Y \rightarrow X$
of Bob and Alice such that 
$g(gr(\phi)) \cap g(gr(\psi)) = \emptyset$. 
Then, we can partition  $\Omega$  into two sets of outcomes 
$\Omega_A$  and  $\Omega_B$ 
(winning for Alice and Bob, respectively) 
in such a way that  
$g(gr(\phi)) \subseteq \Omega_B$  and 
$g(gr(\psi)) \subseteq \Omega_A$.  
(Note that for tight  $g$  this would not be possible.) 
Then,  $-1 = maxmin < min max = 1$  
in the obtained $\pm 1$ game $(g;\Omega_A,\Omega_B)$ 
and, hence, it has no saddle point.  

\smallskip

The inverse implication  $(iii) \Leftarrow (iv)$, 
(as well as   $(ii) \Leftarrow (iv)$, which looks stronger)  
are proven similarly; see \cite{EF70,Gur73}. 
Assume that a zero-sum game  $(g;u,w)$  has no saddle point. 
Then, (\ref{eqMMMM}) fails and maxmin $<$ minmax. 
Consider arbitrary best response strategies  
$\phi : X \rightarrow Y$  and  $\psi : Y \rightarrow X$
of Bob and Alice, respectively. 
Obviously, 
$g(gr(\phi)) \cap g(gr(\psi)) = \emptyset$ and,   
thus, $g$  is not tight. 

\smallskip  

The last claim means that 
a tight game form is  SP-solvable. 
Moreover, it has a simple SP situation 
in minimal strategies \cite{Gur89}.   

\smallskip 

To finish the proof of the theorem it only remains to show 
implication $(i) \Leftarrow (iv)$, that is, 
tightness implies Nash-solvability.
First, this was done in \cite{Gur75}, 
then, with more details, in \cite{Gur89}. 
Several  different proofs appeared later \cite{BBM90,BG03,DS91,Gur18,GK18}.
In the next section we suggest a new 
(and shortest) proof 
based on an important general property of dual multi-hypergraphs. 

\section{Lexicographical theorem for dual multi-hypergraphs} 
\label{s1}
\subsection{Summary} 
\label{ss10}
Let   $\cA$  and  $\cB$  be 
an arbitrary pair of finite dual multi-hypergraphs 
on a common ground set $\Omega$.
Each of them may have embedded or equal edges.
An edge is called {\em containment minimal} 
(or just {\em minimal}, for short) 
if it is not a strict superset of another edge. 
(Note that minimal edges may still be equal.) 
If  $\cA$  and  $\cB$ are dual then      
\begin{enumerate}
    \item[(j)]
    $A \cap B \neq \emptyset$  for  every pair  $A \in \cA$  and  $B \in \cB$;
    \item[(jj)]
if  $A$  is minimal then for every  $\omega \in A$  there exists a 
(minimal) $B \in \cB$  such that  $A \cap B = \{\omega\}$.

We will extend claim (jj) as follows. A linear order $\succ$  over $\Omega$
uniquely defines a lexicographic order $\succ_\ell$  
 over the power set   $2^\Omega$.  
 \item[(jjj-A)] 
 Let  $A$  be a lexicographically maximal (lexmax) edge of  $\cA$. Then,
 edge  $A$  is minimal in $\cA$  and for every  $\omega \in A$  there exists a 
 (minimal) edge  $B \in \cB$  such that  
 $A \cap B = \{\omega\}$  and  $\omega \succeq \omega'$  for each  $\omega' \in B$. 
\end{enumerate}

By swapping  $A, \cA$  and $B, \cB$, we obtain the dual statement (jjj-B). 

These two statements form the {\em lexicographical theorem} for dual multi-hepergraphs. 
To formulate it accurately, we will need a few definitions. 

\subsection{Lexicographical orders over the subsets.}
\label{ss11}
A linear order  $\succ$  over a set $\Omega$  
uniquely determines a lexicographical order $\succ_\ell$
over the power set  $2^\Omega$ (of all subsets of  $\Omega$)  as follows.
Roughly speaking, the more small elements are out of a set - the better it is.
In particular, $\Omega' \succ_\ell  \Omega''$  
whenever  $\Omega' \subset \Omega''$ and, hence,
the empty set  $\emptyset \subset \Omega$  is the best in $2^\Omega$. 

\begin{remark}
Also  $\{\omega'\} \succ_\ell \{\omega', \omega''\}$  
for any $\omega', \omega'' \in \Omega$  and order $\succ$,   
although set  $\{\omega', \omega''\}$  
gives a chance for a better outcome $\omega''$  if  $\omega' \prec \omega''$; 
see game form  $g_6$ in Figure 1 and  subsection \ref{ss22} for more detail.
\end{remark}

More precisely, to compare  two arbitrary subsets  
$\Omega', \Omega'' \subseteq \Omega$  
consider their symmetric difference
$\Delta = (\Omega' \setminus \Omega'') \cup (\Omega'' \setminus \Omega')$.
Clearly,  $\Delta \neq \emptyset$  
if and only if sets $\Omega'$ and $\Omega''$ are distinct.
Let  $\omega$  be the minimum with respect to  
$\succ$  element in  $\Delta$.
If  $\omega \in  (\Omega' \setminus \Omega'')$ then  $\Omega'' \succ_\ell \Omega'$;
if  $\omega \in  (\Omega'' \setminus \Omega')$ then  $\Omega' \succ_\ell \Omega''$.

We can reformulate this definition equivalently as follows.
Without loss of generality 
(wlog), set $\Omega = \{\omega_1, \dots, \omega_p\}$ and 
assume that  $\omega_1 \prec \dots  \prec \omega_p$;   
assign the negative weight  $w(\omega_i) = -2^{k-i}$  to every $\omega_i \in \Omega$, 
and set  $w(S) = \sum_{\omega \in S} w(\omega)$   for each subset  $S \subseteq \Omega$. 
Then, $\Omega' \succ_\ell \Omega''$  if and only if  $w(\Omega') > w(\Omega'')$.

Denote by $supp(S)$  the $0,1$-vector $(s_1, \dots, s_k)$, 
where  $s_i = 1$  if and only if  $\omega_i \in S$. 
Then obviously, $\Omega' \succ_\ell \Omega''$  if and only if 
$supp(\Omega')$ is less than  $supp(\Omega'')$ 
in the standard lexicographical order.  

\subsection{Dual multi-hypergraphs}  
Two finite multi-hypergraphs  $\cA$  and  $\cB$  
on the common ground set  $\Omega$  are called {\em dual}  
if  (j)  holds:  $A \cap B \neq \emptyset$  for  every pair  
$A \in \cA$  and  $B \in \cB$,  and also 

\smallskip 
(jv-A) 
for each  $B^T \subseteq \Omega$  such that 
$B^T \cap B \neq \emptyset$  for  every  $B \in \cB$ 
there exists an  $A \in \cA$  such that  $A \subseteq B^T$. 

\smallskip 

If (j) and (jv-A) both hold we say that $\cA$ is dual to $\cB$  and 
use notation $\cA = \cB^d$. 
Swapping  $A, \cA$  and  $B, \cB$  in (jv-A) we obtain  (jv-B) 
and an equivalent definition of duality, 
that is, (j)  and  (jv-A)  hold  if and only  (j) and  (jv-B)  hold.
In other words, $\cA = \cB^d$  if and only if  $\cB = \cA^d$. 
So we just say that multi-hypergraphs  $\cA$  and $\cB$  are dual.

\begin{remark}
Dual  multi-hypergraphs have numerous applications and  
appear in different areas 
under different names, such as ``clutters" and  ``blockers" \cite{EF70} 
or DNFs and CNFs of monotone Boolean functions \cite{CH11}. 
\end{remark}

\subsection{Lexicographical Theorem} 
\label{ss13}
Claims (j) and (jj) are well-known 
\cite{CH11}. 
Actually, (j) is required by the definition of duality 
and (jj)  is obvious. 
Indeed, if (jj) fails then edge  $A$ cannot be minimal, 
since its proper subset 
$A \setminus \{\omega\}$  would still intersect all  $B \in \cB$.

Our main result is statement (jjj-A). 
Fix an arbitrary order $\succ$ over $\Omega$  and 
find a lexmax edge  $A^L \in \cA$, that is, one maximal 
with respect to the lexicographical order  $\succ_\ell$ over $2^\Omega$.  
Note that such  $A^L$  may be not unique but all lexmax edges are  equal. 
The lexicographic theorem is formulated as follows: 

\begin{theorem}
\label{t-lex}
A lexmax edge  $A^L$  is minimal in $\cA$.
Furthermore, for every $\omega^* \in A^L$  there exists a  (minimal) edge  
$B^M \in \cB$  such that  $A^L \cap B^M = \{\omega^*\}$  
and  $\omega^* \succ \omega$  for each  $\omega \in B^M \setminus \{\omega^*\}$.
\end{theorem}

\proof  
A lexmax edge must be minimal,
since a set is strictly less than any its proper subset in order $\succ_\ell$.

Assume for contradiction that there exists an  $\omega^* \in A^L$  such that
for every  (minimal)   $B \in \cB$  satisfying 
(jj), $B \cap A^L = \{\omega^*\}$, 
there exists  an  $\omega \in B$  such that  $\omega \succ \omega^*$. 
Clearly, this assumption holds for every $B^0 \in \cB$ 
if it holds for each minimal  $B^0 \in \cB$.
Let us show that it contradicts the lexmaximality of  $A^L$.
To do so partition all edges  $B \in \cB$  into two types:
\begin{enumerate}[(a)]
    \item there is an  $\omega \in  B \cap A^L$  distinct from   $\omega^*$; 
    \item $B \cap A^L = \{\omega^*\}$. 
\end{enumerate}

In case (b), by our assumption,  
there is an  $\omega \in B$  such that  $\omega \succ \omega^*$.

In both cases, (a) and (b),   
choose the specified  $\omega$  from $B$, 
thus, getting a transversal  $B^T$.  By (jv-A), 
there exists an  $A \in \cA$  such that $A \subseteq B^T$ and, 
hence, $A \succeq_\ell B^T$. 
Furthermore, by construction, $B^T \succ_\ell A^L$. 
Indeed,  $\omega^* \not\in B^T$  and it is replaced in  $B^T$  
by some larger elements, $\omega \succ \omega^*$, in case (b), 
while all other elements of  $B^T$, if any, 
belong to  $A^L \setminus \{\omega^*\}$, according to case (a). 

Thus, by transitivity,  $A  \succ_\ell A^L$.  
Yet, by assumption of the theorem, 
$A^L$  is a lexmax edge of $\cA$, which is a contradiction.
\qed

\subsection{Sperner hypergraphs}  
\label{ss12} 
A multi-hypergraph is called {\em Sperner} 
if no two of its distinct edges contain one another;
in particular, they cannot be equal. 
In this case,
we have a hypergraph rather than multi-hypergraph. 
For a multi-hypergraph there exists a unique 
dual Sperner hypergraph. 
If  $\cA$  and  $\cB$  are dual and Sperner then 
$\cA^{dd} = \cA$  and  $\cB^{dd} = \cB$; furthermore 
$\cup_{A \in \cA} A = \cup_{B \in \cB} B = \Omega$.
In general, for multi-hypergraphs, 
$\cup_{A \in \cA} A$  and  $\cup_{B \in \cB} B$  
may be different subsets of  $\Omega$. 

\begin{remark} 
\label{r-Sperner} 
Here we assume that the reader is familiar with basic notions
related to monotone Boolean functions, in particular,  
with DNFs and duality. 
An introduction can be found in \cite [Sections 1, 3 and 4]{CH11}.

It is well known \cite{CH11} that 
(dual) multi-hypergraphs are in one-to-one correspondence with (dual) monotone DNFs:  
(prime) implicants of the latter correspond to 
(minimal) edges of the former.
Furthermore, Sperner hypergraphs 
correspond to irredundant DNFs.
However, we do not restrict ourselves to this case. 
Although the lexicographical theorem  would not lose much but 
its applications to Nash-solvability would.
\end{remark} 

\section{Determining edges  $A^L$ and $B^M$ 
of Theorem \ref{t-lex} in polynomial time.}
\label{ss14}
\subsection{Preliminaries}
Edges  $A$  and  $B$  mentioned in (jjj-A)  
can be found in polynomial time. 
The problem is trivial when  multi-hypergraphs
$\cA$  and  $\cB$  are given explicitly.
We will solve it when only  $\cA$  is given, and not explicitly, 
but by a polynomial containment oracle. 
For an arbitrary subset  $\Omega_A \subseteq \Omega$  
this oracle answers in polynomial time the question 
$Q(\cA, \Omega_A)$: whether  $\Omega_A$  contains an edge $A \in \cA$.

By duality of $\cA$  and $\cB$, we have  
$A \not \subseteq \Omega_A$  for all  $A \in \cA$  if and only if
$B \subseteq \Omega_B = \Omega \setminus \Omega_A$  for some  $B \in \cB$.
In other words, question $Q(\cA, \Omega_A)$  
is answered in the negative if and only if
$Q(\cB, \Omega_B)$  is answered in the positive. 
Thus, we do not need two separate oracles for  $\cA$ and $\cB$; 
it is sufficient to have one, say, for $\cA$.

\subsection{Determining a lexmax edge $\cA^L$ in polynomial time.} 
Recall that multi-hypergraph  $\cA$  may contain several lexmax edges  $A^L$,
but they are all equal. 
Fix an arbitrary linear order  $\succ$  over  $\Omega$.
Wlog we can assume that  $\Omega = \{\omega_1, \dots, \omega_p\}$  
and  $\omega_1 \prec \dots \prec \omega_p$.

\smallskip 

Step 1: 
Consider  $\Omega^1_t = \{\omega_t, \dots, \omega_p\}$ and, 
by asking question $Q(\cA, \Omega^1_t)$ for  
$t = 1, \dots, p$, find the maximum  $t_1$  for which the answer is still positive.
Then,  $\omega_{t_1}$  belongs to  $\cA^L$,  
while  $\omega_1, \dots, \omega_{t_1 - 1}$  do not.  

\smallskip 

Step 2: 
Consider  $\Omega^2_t = \{\omega_{t_1}, \omega_{t_1 + t}, \dots, \omega_p\}$  
and, by asking question $Q(\cA, \Omega^2_t)$ for  
$t = 1, \dots, p - t_1$, find the maximum  $t_2$  
for which the answer is still positive. 
Then,  $\omega_{t_1}, \omega_{t_1 + t_2} \in \cA^L$,  
while  $\omega_t \not \in \cA^L$  for any other  $t < t_1 + t_2$. 

\smallskip 

Step 3: 
Consider  
$\Omega^3_t = \{\omega_{t_1}, \omega_{t_1 + t_2}, \omega_{t_1 + t_2 + t}, \dots, \omega_p\}$  
and, by asking question $Q(\cA, \Omega^3_t)$ for  
$t = 1, \dots, p - (t_1 + t_2)$, find the maximum  $t_3$  
for which the answer is still positive. 
Then,  $\omega_{t_1}, \omega_{t_1 + t_2}, \omega_{t_1 + t_2 + t_3} \in \cA^L$, 
while  $\omega_t \not \in \cA^L$  for any other  $t < t_1 + t_2 + t_3$;  etc. 

\smallskip 

This procedure will produce a lexmax edge $A^L$ 
in at most  $p$  polynomial iterations. 
Note that on each step  $i$ 
we can speed up the search of $t_i$  by applying dichotomy. 

\medskip 

\subsection{Determining an edge $B^M$  from Theorem \ref{t-lex}.}
First, find a lexmax edge  $A^L \in \cA$  
and choose an arbitrary  $\omega^* \in A^L$. 
We look for an edge  $B^M \in \cB$  such that  
$A^L \cap B^M = \{\omega^*\}$  and  $\omega^* \succ \omega$  
for every  $\omega \in B^M \setminus \{\omega^*\}$. 
In other  ``words", 

$$B^M \subseteq  \Omega_B = 
\Omega  \setminus [(A^L  \setminus \{\omega^*\}) \cup \{\omega \mid \omega \succ \omega^*\}].$$ 

By Theorem \ref{t-lex}, such  $B^M$  exists and, hence, 
the oracle answers  $Q(\cB, \Omega_B)$ in the positive, 
or equivalently,  $Q(\cA, \Omega \setminus \Omega_B)$  in the negative. 
We could take any $B^M \in \cB$  such that  $B^M \subseteq \Omega_B$. 
Yet, multi-hypergraph  $\cB$  is not given explicitly. 
To get  $B^M$  we need ``to minimize"  $\Omega_B$. 
To do so, let us delete its elements one by one 
in some order until we obtain  
a minimum set  $\Omega_B^*$  for which the answer to 
$Q(\cA, \Omega \setminus \Omega_B^*)$  is still negative, that is,   
answers  to  $Q(\cA, \Omega \setminus (\Omega_B^* \setminus \{\omega\})$  
become positive for every $\omega \in \Omega_B^*$. 
Then, we set  $B^M = \Omega_B^*$. 
Again we can speed up the procedure by applying dichotomy.

Note that the above reduction procedure may be not unique, since 
we can eliminate elements of  $\Omega \setminus \Omega_B$  in an arbitrary order.
Thus, in contrast to  $A^L$, there may be 
several not equal edges  $B^M$  
satisfying all conditions of Theorem~\ref{t-lex}. 

\section{Lexicographically safe NE in games with tight game forms} 
\label{s2}
\subsection{Summary} 
\label{ss20}
First, we apply Theorem \ref{t-lex} 
to finish the proof of Theorem \ref{t-NS}. 
It remains to show that $(i) \Leftarrow (iv)$, 
that is, tightness implies Nash-solvability. 
In other words, a game  $(g;u,w)$  
has a NE for any payoffs  $u$ and $w$ 
whenever game form  $g$  is tight. 
The proof is constructive: we will obtain two 
special types of NE. 

Given  $g$  and  $u$,  choose a lexmax strategy 
$x \in X$  of Alice. 
By Theorem~\ref{t-lex}, there is a strategy 
$y \in Y$  of Bob  such that  $(x,y)$  is a NE. 
By definition,  $y$  must be  a best response to  $x$  
such that  $x$  is also a best response to  $y$. 
By Theorem \ref{t-lex}, 
the obtain situation $(x,y)$ is simple  
and both strategies, $x$ and $y$ are minimal. 
More precisely, $x$  must be minimal, while 
$y$  can be chosen minimal. 
These NE will be called {\em lexsafe NE of Alice} and 
the set of these NE will denoted by  NE-A.
Similarly, we define a set NE-B of Bob's {\em lexsafe NE}.

\begin{remark}
We assume that both players are maximizers and  
adjective ``lexsafe" can be replaced by ``lexmax". 
If both players are minimizers then it can be replaced by ``lexmin".
In the zero-sum case 
Alice is the maximizer, while Bob is the minimizer.
Flexible term lexsafe may replace both, lexmax or lexmin.
\end{remark}

Results of Section \ref{ss14} 
provide a polynomial algorithm determining  
at least one NE from NE-A and at least one from NE-B 
(which may coincide)  
in a given game $(g;u,w)$  with a tight game form  $g$. 
This is trivial when  $g$  is given explicitly. 
Yet, the algorithm works when 
one of two multi-hypergraphs  $\cA(g)$ or $\cB(g)$ 
is given by a polynomial containment oracle. 

\subsection{Lexicographically safe strategies of players}
\label{ss22}
Given $g$ and preference $\succ_A$ of Alice, 
let us introduce the lexicographical 
pre-order over Alice's strategies $x \in X$  as follows. 
Consider lexicographical order  $\succ_A^\ell$  over $2^\Omega$ 
defined by the linear order $\succ_A$  over  $\Omega$. 
The larger is the support  $g(x) \subseteq \Omega$  
in order $\succ_A^\ell$,
the safer is strategy  $x$  for Alice,  
while strategies with the same support are equally safe.
Alice's strategies that maximize 
support  $g(x)$  in order $\succ_A^\ell$
will be called her lexmax (or  lexsafe) strategies.

In particular, all lexmax strategies have the same support. 

Furthermore, a lexsafe strategy is minimal.
Indeed, $x$  is safer than $x'$  whenever $g(x) \subset  g(x')$ 
and containment is strict. 

Note also that Alice's lexsafe strategies are
defined by $g$ and $\succ_A$, while Bob's preference $\succ_B$ is irrelevant.  
Alice may be even unaware of it, which is important for applications.

\smallskip 

Similarly, using $y \in Y$ and  $\succ_B$ 
instead of  $x$  and  $\succ_A$, 
we define Bob's lexsafe strategies. 
Respectively, they depend only on  $g$  and  $\succ_B$, 
while  $\succ_A$  is irrelevant. 

\medskip 

The concept of a lexsafe strategy
can be viewed as a refinement 
of the classical concept of a safe (maxmin) strategy.
The latter optimizes the worst case scenario outcome,
while lexsafe strategies optimize the whole set of outcomes
in the lexicographical order defined above. 

Thus, lexsafe strategies are safest, 
but  sometimes may be not rational. 
For example, let  
$g(x) = \{\omega\}, \; g(x') = \{\omega, \omega'\}$  
and $\omega \prec_A \omega'$.
Then  $x \succ_A  x'$, although 
strategy  $x'$  is  better for Alice than  $x$. 
Indeed, $x'$  gives her a chance to obtain 
the better outcome  $\omega'$, 
while  $x$  excludes $\omega'$ and ensures $\omega$; see Remark~5.

Consider a preference over  $\Omega$  
such that outcomes $\omega, \omega' \in \Omega$ 
are, respectively, the worst and the best outcomes 
for both Alice and Bob simultaneously. 
Consider a game form 
$g : X \times Y \rightarrow \Omega$  
having two strategies  $x^* \in X$ and $y^* \in Y$  such that 
$g(x,y) = \omega$  if and only if 
$x = x^*$  or  $y = y^*$ 
and  $g(x,y) = \omega'$  otherwise. 
Then, $x^*$  and  $y^*$  are 
the only lexsafe strategies of Alice and Bob;
furthermore, situation  $(x^*, y^*)$  is a unique lexsafe  NE, 
but outcome  $g(x^*, y^*) = \omega$  is 
worse than  $\omega'$  for both players.  
One can view this as a price of stability. 
For comparison, recall that NE may be not Pareto optimal. 

\subsection{Lexsafe Nash equilibria in games with tight game forms}  
\label{ss23}
Recall that game form  
$g : X \times Y \rightarrow \Omega$  is tight 
if and only if its hypergraphs  
$\cA = \cA(g) = \{g(x) \mid x \in X\}$  and  $\cB = \cB(g) = \{g(y) \mid y \in Y\}$ 
are dual. 

Given a game $(g;\succ_A,\succ_B)$  with a tight game form  $g$, 
choose any lexsafe strategy  $x^L$  of Alice. 
By Theorem \ref{t-lex}  it is minimal. 
Let us show that there exists a strategy  $y^M$  of Bob such that 
$(x^L, y^M)$  is a NE. 
By definition, $y^M$  is a best response to  $x^L$, 
that is, $g(x^L,y^M) \succeq_B  g(x^L,y)$  for any  $y \in Y$. 
(Note, however, that the preference is not strict, 
because for some  $y$  two outcomes may coincide: $g(x^L,y^M) = g(x^L,y)$.)  
Let us apply Theorem  \ref{t-lex}  setting 
$$g(x^L) = A^L, \; g(y^M) = B^M, g(x^L,y^M) = \omega^*,$$ 
and conclude that there exists a (minimal) strategy $y^M$  such that 
$x^L$, in its turn, is a best response to  $y^M$. 
Thus, $(x^L,y^M)$  is a NE. Theorem \ref{t-NS} is proven. 

\smallskip 

Moreover, we can strengthen it  
summarizing remarkable properties of the obtained NE. 
Recall that in Theorem \ref{t-lex} 
both edges  $A^L$  and  $B^M$  are minimal and  
$A^L \cap B^M = \{\omega^*\}$.
Hence, for the obtained NE  $(x^L,y^M)$ 
both strategies $x^L$  and  $y^M$  are minimal  and 
situation   $(x^L, y^M)$  is simple, that is, 
$g(x^L) \cap g(y^M) = \{\omega^*\}$; see \cite{Gur89}. 
More precisely, $X^L$ must be minimal, 
while $Y^L$ can be chosen minimal. 

Denote by  $X^L$  the set of all lexmax strategies of Alice. 
By definition, they all have the same support.  
Let us fix  $x^L \in X^L$  and denote by  $Y^M(x^L)$ 
the set of all 
Bob's  best responses to  $x^L$. 
In fact,  $Y^M(x^L)$  does not depend on  $x^L$  provided   $x^L \in X^L$. 
Indeed, set  $g(x^L)$  is unique, that is, 
the same for all  $x^L \in X_L$  and 
$g(x^L) \cap g(y^M) = \omega^*$   for all  $y^M \in Y^M(x'^L)$ 
and for all  $x'^L \in X^L$. 
Hence,  $y^M(x^L)$  is a best response of Bob to 
each Alice's lexmax strategy. 
Denote  by  $y^M$  the set of all such best responses. 


Thus, we obtain 
$X^L \subseteq X$  and  $Y^M \subseteq Y$  such that 
for any pair  $x^L \in X^L$  and  $y^M \in Y^M$  
situation  $(x^L, y^M)$  is simple,  
$g(x^L, y^M) = \{\omega^*\}$,  and   $(x^L, y^M)$  is a NE, 
because  $X^L$  is a best response to $Y^M$  and vice versa. 

In other words, the direct product  
NE-A $= (x^L \times y^M) \subseteq X \times Y$ 
consists of simple NE situations 
corresponding to the same outcome $\omega^* \in \Omega$. 
Furthermore, all strategy of $X^L$  are lexsafe and, 
hence, minimal, while  $Y^M$  contains minimal strategies. 
We will call  NE-A  the {\em box of Alice's lexsafe equilibria}. 

\smallskip

By construction, 
$X^L$  depends only on Alice's  preference  $\succ_A$, while 
Bob's preference  $\succ_B$  is irrelevant and  
Alice may be just unaware of it, which is important for applications. 
In contrast, $Y^M$  is a set of some 
(special) Bob's best responses to  $X^L$, 
which are the same for all  $x^L \in X^L$. 

\medskip 

Swapping the players, we obtain the 
{\em box of Bob's lexsafe equilibria} 
NE-B $= (x^M \times y^L) \subseteq X \times Y$ 
with similar properties. 
Thus, we can strengthen Theorem \ref{t-NS}  as follows: 

\begin{theorem}
\label{t-NS+} 
Every game  $(g;\succ_A,\succ_B)$  with a tight game form  $g$
has two non-empty boxes of lexmax equilibria 
NE-A = $X^L \times Y^M$  and  NE-B = $X^M \times Y^L$   
of Alice and Bob satisfying the above properties.
\qed 
\end{theorem}


Boxes  NE-A  and  NE-B  may intersect or even coincide. 
For example, this always happens in the sero-sum case.. 
In this case $X^L$ and $X^M$ are maxmin strategies of  Alice, 
while $Y^M$ and $Y^L$ are minmax strategies of  Bob. 
More detail can be found 
in the first arXiv version of this paper \cite{GN21}. 
NE-A and NE-B may be equal in the non-zero-sum case too. 
For example a game may have a unique NE. 

\subsection{A pair of lexsafe strategies of Alice and Bob may be not a NE} 
\label{ss08a}
For example, consider tight game form  $g_1$  in Figure 1. 
Define preferences  $\succ_A$  and  $\succ_B$  such that
$\omega_2 \succ_A \omega_1 \succ_A \omega_3$  and
$\omega_2 \succ_B  \omega_3$.  
It is easy to  verify that
$x_1$  and  $y_1$  are lexsafe strategies of Alice  and Bob, respectively.
Yet, situation  $(x_1,y_1)$   is not an NE.
Alice can improve her result   $g_1(x_1,y_1) = \omega_1$  
by switching to  $x_2$  and getting  $g(x_2,y_1) = \omega_2$.
Thus, two lexsafe strategies, of Alice and Bob, do not form an NE.
However, sets  NE-A and NE-B are not empty, 
in accordance with Theorem \ref{t-NS+}: 
NE-A $= \{(x_2,y_1)\}$  and  NE-B $= \{(x_1,y_2)\}$. 
The corresponding NE outcomes are $\omega_1$ and $\omega_2$, respectively.

Note that $\omega_2$ is the best outcome for both players 
if  $\omega_2) \succ_B \omega_1$. 
In this case NE-B is not Pareto-optimal.

\begin{remark}
One could conjecture that
each player prefers lexsafe NE of the opponent to his/her own.
Such result would be similar to the analogous  one
from the matching theory; see, for example, \cite{GI89}.
There are two types of stable matchings
given by the Gale-Shapley algorithm \cite{GS62},
depending on  men propose to women or  vice versa.
Yet, this conjecture is disproved by the above example.
\end{remark}

\section{Computing lexsafe NE in polynomial time}
\label{s3}
If game form  $g : X \times Y \rightarrow \Omega$
is given explicitly then to find all its NE is simple:
one can just consider all situations  $(x,y) \in X \times Y$
one by one verifying Nash's definition for each of them.
Yet, in applications $g$  is frequently given by an oracle $\cO$ such that
size of  $g$  is exponential in size  $|\cO|$  of this oracle.
Then, the straightforward search for NE  suggested above becomes not efficient.
Four such oracles will be considered in the next section.
The following three properties of oracle  $\cO$
will allow us to construct an algorithm computing two lexsafe NE
(from NE-A and NE-B, respectively) for a given game
$(g;\succ_A,\succ_B)$  with tight game form  $g = g(\cO)$  realized by $\cO$,
in time polynomial in $|\cO|$.

\begin{enumerate}[(I)]
\item  Oracle  $\cO$  contains explicitly 
all outcomes  $\Omega$  of  $g$.
\newline
(Yet, strategies  $x \in X$  and  $y \in Y$  are implicit in  $\cO$;
moreover, $|X|$  and  $|Y|$  may be exponential in $|\cO|$.)

\item  The game  form  $g = g(\cO)$  defined by  $\cO$  is tight.

\item  Every $\pm 1$ game  $(g(\cO);\Omega_A,\Omega_B)$  can be solved
in time polynomial in  $|\cO|$.
\end{enumerate}

Requirement (III) needs a discussion.
By tightness of  $g$,  
exactly one of the following two options holds:

\begin{enumerate}[(a)]
\item  there exists $x \in X$  with  $g(x) \subseteq \Omega_A$
 (Alice wins);

\item  there exists $y \in Y$  with  $g(y) \subseteq \Omega_B$
 (Bob wins).
\end{enumerate}

Note that (a)  (respectively, (b))  holds if and only if 
the monotone Boolean function corresponding 
to multi-hypergraph  $\cA(g)$  (respectively, $\cB(g)$ 
takes value $1$; see Remark \ref{r-Sperner}.

\smallskip 

 To solve a $\pm 1$ game we determine which option,  
 (a) or (b),  holds and output a winning strategy,
 $x$  or $y$, respectively.

Note that it is possible to output a
{\em minimal} winning strategy whenever (III) holds.
Indeed, suppose Alice wins and we output her winning strategy  $x$,
with  $g(x) \subseteq  \Omega_A$.
Reduce  $\Omega_A$  by one outcome $\omega$ by  moving it to  $\Omega_B$, 
solve the obtained $\pm 1$ game, and 
repeat the procedure for all  $\omega \in \Omega_A$.
If Bob wins in all obtained games 
then  $x$  is already minimal.
Otherwise we can move an outcome $\omega$ from $\Omega_A$  to $\Omega_B$ and Alice still wins.
Repeating, we obtain a minimal winning strategy of Alice
(in the original game) in at most $|\Omega_A|$  steps.
We can speed up the above procedure using dichotomy. 
The same works for Bob.

\medskip 

Theorems 1--3 immediately imply the following statement. 

\begin{theorem}
\label{t2}
Given an oracle  $\cO$  satisfying requirements 
(I,II,III), a lexsafe NE of Alice (of Bob) exists and
can be computed in time polynomial in $|\cO|$.
\qed
\end{theorem}

\section{Examples of oracles}
\label{s4}
\subsection{Summary}
Here we consider four types of oracles known in the literature 
and verify that all four satisfy requirements (I, II, III).

\medskip 

In Section 8.3 we consider game forms corresponding 
to positional (graphical) game structures with perfect information, 
due to which Nash-solvabilty holds even in the $n$-person case. 
Yet, we substantially extend this class of game forms 
by modifying the set of outcomes. 
The standard approach assumes that 
the set of outcomes $\Omega$ is formed 
by the terminal vertices of the input directed graph  $\Gamma$. 
Yet, Nash-solvability still holds 
if we extend $\Omega$ by redefining it as the set of all 
strongly connected components of  $\Gamma$.
Yet, in this case Nash-solvability holds 
only if we restrict ourselves to 2-person games. 

In Section 8.4 we introduce so-called Jordan game forms 
in which  Alice and Bob connect two pairs of opposite 
sides of the square. 
In Section 8.5 we consider monotone bargaining 
and in Section 8.6 veto voting schemes. 
In these three examples perfect information is not assumed,  
nevertheless requirements  (I,II,III), 
tightness among them, hold.

Subsections  8.3 - 8.6  can be read in an arbitrary order. 
 
\subsection{Game forms and game correspondences}
\label{ss40A} 
A {\em game correspondence}  is defined as an arbitrary mapping
$G : X \times Y \rightarrow  2^\Omega \setminus \{\emptyset\}$,
that is,  $G$  assigns a non-empty subset of outcomes to each situation.

Given $G$, define a game form  $g \in G$, choosing
an arbitrary outcome $g(x,y) \in G(x,y)$  for each situation  $(x,y)$.
Conversely, given a game form  $g : X \times Y \rightarrow \Omega$,
define a game correspondence  $G$  setting  $G(x,y) = g(x)\cap g(y)$.
Then, obviously,  $g \in G$.

By property (jj) of Section \ref{s1}, 
if at least one  $g^* \in  G$  is tight
then all $g \in G$  are tight.
In this case  $G$  is called {\em tight} too.
Moreover, all  $g \in G$  have the same Sperner 
reduced dual hypergraphs $\cA^0(g)$ and $\cB^0(g)$,
same simple situations, and for any 
$u$ and $w$, the same sets of simple situations in NE-A and NE-B.

\subsection{Deterministic graphical multi-stage game structures}
\label{ss40}
Let $\Gamma = (V,E)$ be a directed graph (digraph) whose vertices
and arcs are interpreted as positions and moves, respectively.
Denote by  $V_T$  the set of terminal positions 
(of out-degree zero) 
and by  $V_A, V_B$  the sets of  positions of positive out-degree 
controlled by Alice and Bob, respectively.
We assume that  $V = V_A  \cup  V_B \cup V_T$  is a partition of  $V$.

A strategy $x \in X$  of Alice
(respectively, $y \in Y$  of Bob) is a mapping that
assigns to each position  $v \in V_A$
(respectively, $v \in V_B$) a move from this position.
An initial position  $v_0 \in V_A \cup V_B $ is fixed.
Each situation $(x,y)$  defines a unique walk in $\Gamma$ 
that begins in  $v_0$  and then 
follows the decisions made by strategies  $x$  and $y$.
This walk $P(x,y)$ is called a {\em play}.
Each play either terminates in  $V_T$  or is infinite.
In the latter case, it forms a ``lasso":
first, an initial path, which may be empty, and
then, a directed cycle (dicycle) repeated infinitely.
(Indeed, since both players are restricted  
to their stationary strategies, 
a move may depend only on the current position but
not on previous positions and/or moves. Hence, 
if a play visits a position twice then  
all further moves will be repeated as well.) 

The (positional structure 
defined above can also be represented in normal form.
We introduce a game form  $g : X \times Y \rightarrow \Omega$,
where, as before, $\Omega$  denotes a set of outcomes.
Yet, there are several ways to define this set.
One is to ``merge" all infinite
plays (lassos) and consider them as a single outcome  $c$,
thus, setting  $\Omega = V_T \cup \{c\}$.
This model was introduced by Washburn \cite{Was90}
and called {\em deterministic graphical game structure} (DGGS).

The following generalization was suggested in \cite{Gur18}.
Digraph  $\Gamma$  is called {\em strongly connected}  if
for any  $v, v' \in V$  there is a directed path from  $v$  to  $v'$
(and, hence, from  $v'$ to $v$, as well).
By this definition, the union of two strongly connected digraphs
with a common vertex is strongly connected.
A vertex-inclusion-maximal strongly connected induced subgraph of $\Gamma$
is called its {\em strongly connected component} (SCC).
In particular, each terminal position $v \in V_T$ is an SCC.
It is both obvious and well-known that
any digraph  $\Gamma = (V, E)$  admits a unique decomposition into SCCs:
$\Gamma^\omega = \Gamma[V^\omega] = (V^\omega, E^\omega)$  for $\omega \in \Omega$, where
$\Omega$  is a set of indices.
Furthermore, partition  $V = \cup_{\omega \in \Omega} V^\omega$  can be constructed in time linear
in the  size of  $\Gamma$, that is, in $(|V| + |E|)$.

Partitioning into SCCs has numerous applications; 
see \cite{Sha81,Tar72} for more details.
One more application was suggested in \cite{Gur18}.
For each  $\omega \in \Omega$, contract the SCC  $\Gamma^\omega$  into a single vertex  $v^\omega$.
Then, all edges of  $E^\omega$ (including loops) disappear and
we obtain an acyclic digraph  $\Gamma^* = (\Omega, E^*)$.
Set  $\Omega$  can be treated as the set of outcomes.
Each situation  $(x,y)$ uniquely defines a play $P = P(x,y)$.
This play either comes to a terminal  $v \in V_T$  or forms a lasso.
The cycle of this lasso is contained in an SCC  $\omega$  of $\Gamma$.
Each terminal is an SCC as well.
In both cases an SCC  $\omega \in \Omega$ is assigned to the play  $P(x,y)$.
Thus, we obtain a game form $g : X \times Y \rightarrow \Omega$,
which is the normal form of the
{\em multi-stage DGGS} (MSDGGS) defined by $\Gamma$.

An SCC is called {\em transient} if it is not a terminal
and contains no dicycles. 
Obviously, a transient SCC consists of a single vertex   
and no play can result in it. Thus, it is not an outcome.
For example, $\Omega = V_T$  
in any acyclic digraph, while each remaining SCC is transient.  

\begin{proposition}
In both cases, DGGS and MSDGGS, the corresponding oracles
satisfy requirements  (I, II, III).
\end{proposition}

\proof 
Indeed, (I) holds since the  
SCCs, of a given digraph $\Gamma$ 
can be generated in time linear in the size of $\Gamma$.

Both requirements, (II) and (III), 
for both oracles, DGGS and MSDGGS, can be verified simultaneously.
Consider the corresponding game forms  $g'$  and  $g$  and note
that  $g'$  is obtained from  $g$  by merging some outcomes.
Namely, all outcomes corresponding to the non-terminal SCCs
are replaced by a single outcome $c$.
It is both obvious and well-known that merging outcomes respects tightness.
Hence, it is enough to verify  (II) and (III) for MSDGGSs,

By Theorem \ref{t-NS}, to verify (II) 
it is sufficient to prove $\pm 1$ solvability.
For DGGS it was done in \cite{Was90}; see also
\cite[Section 3]{BG03}, \cite{AHMS08}, \cite[Section 12]{BGMS07}.
This result was extended to MSDGGS in \cite{Gur18}.
Furthermore, all proofs in \cite{Gur18} were constructive,
the corresponding $\pm 1$ games were solved in time
polynomial in the size of  $\Gamma$, which implies (III).

\medskip

For reader's convenience, we briefly sketch here 
the proof of (II,III) from \cite{Gur18}.
Consider a $\pm 1$ game $(g;\Omega_A,\Omega_B)$ with game form  $g = g(\cO)$
generated by a MSDGGS oracle $\cO$.
We would like to apply Backward Induction,
yet, digraph $\Gamma$  may have dicycles.
So we modify Backward Induction to make it work in presence of dicycles.
Recall that $\Omega$  is the set of SCCs of  $\Gamma$ and  
$\Gamma^* = (\Omega, E^*)$  is acyclic.
Consider an SCC  $\Gamma' = (V',E')$  in  $\Gamma$
that is not terminal, but each move  $(v',v)$
from a position  $v' \in V'$
either ends in a terminal  $v \in V_T$, or
stays in  $\Gamma'$, that is, $v, v' \in V'$. 
Obviously, such a SCC exists. 
Note that it may be transient. 
In this case the standard Backward Induction is applicable. 

Suppose that $\Gamma'$  is not transient,
in other words, it contains a dicycle.
Wlog we can assume that $\omega \in \Omega_A$, that is,
Alice wins if the play cycles in $\Gamma'$.
Then, Bob wins in a position $v' \in V'$  if and only if
he can force the play to terminate in  $\Omega_B$, while  
Alice wins in all other positions of $V'$.
Note that it is not necessary for Alice 
to force the play to come to a terminal from $\Omega_A$,  
if the play cycles in $\Gamma'$ Alice wins as well.
Thus, every position of  $\Gamma'$  belongs  
either to $\Omega_A$  or to $\Omega_B$. 
We make all these positions terminal, 
by eliminating all edges  $E'$  of  $\Gamma'$, and repeat
until the initial position $v_0$  of $\Gamma$  is evaluated.

This procedure proves solvability of game form  $g = g(\cO)$ 
(which is equivalent to its tightness (II), by Theorem 1),   
moreover, a $\pm 1$ game $(g;\Omega_A,\Omega_B)$  
is solved in time linear in the size of  $\cO = \Gamma$ 
(which is (III)).
\qed

\subsubsection*{Acyclic deterministic graphical game structures}
A game form  is called \emph{rectangular} if all its situations are simple.
It is shown in  \cite{Gur82}  that a game form  $g$ is generated
by a DGGS whose graph is a tree
if and only if  $g$  is tight and rectangular.
Two examples,  $\Gamma_1$  and  $\Gamma_2$  
generating game tight rectangular game forms  $g_1$  and $g_2$ 
are given in Figure \ref{f2}; see also Figure~\ref{f1}.
More examples can be found in \cite[Section 3]{Gur09}, where
the above characterization is extended to the $n$-person case.

\tikzset{ell/.style=
    {ellipse,draw,minimum height=0.2cm, minimum width=0.2cm, inner sep=0.1cm}}

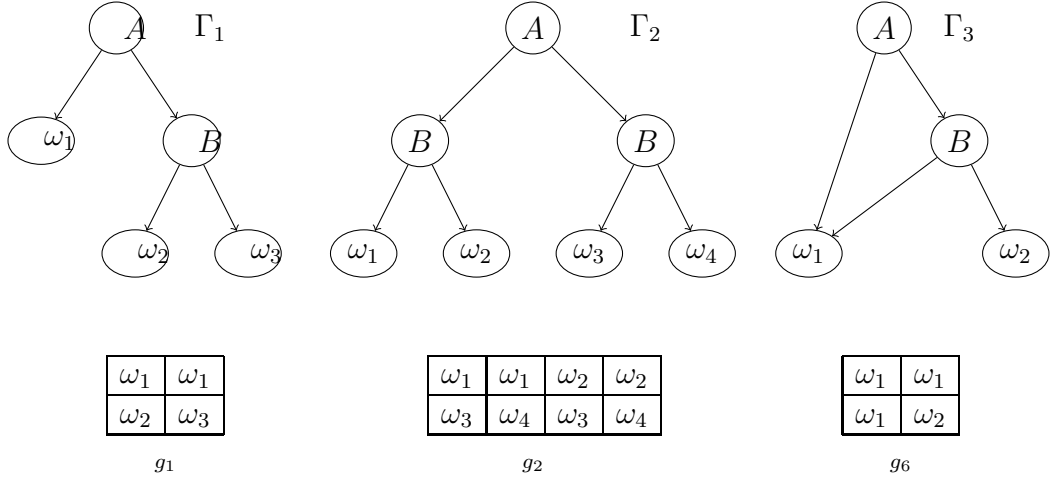
\begin{figure}[t]
\begin{tikzpicture}
\hspace{0.25cm}
\node[ell] (A)at (-14.5,1.5){$A$};
\node[ell] (o1)at (-15.5,0){$\omega_1$};
\node[ell] (B)at (-13.5,0) {$B$};
\node[ell] (o2)at (-14.25,-1.5) {$\omega_2$};
\node[ell] (o3)at (-12.75,-1.5) {$\omega_3$};
\draw [->] (A) to (o1);
\draw [->] (A) to (B);
\draw [->] (B) to (o2);
\draw [->] (B) to (o3);
\node at (-13.5,1.5) {$\Gamma_1$};
\end{tikzpicture}%
\hspace{0.5cm}
\begin{tikzpicture}
\node[ell] (A_)at (-1,1.5){$A$};
\node[ell] (B_1)at (-2.5,0){$B$};
\node[ell] (B_2)at (0.5,0) {$B$};
\node[ell] (o_1)at (-3.25,-1.5) {$\omega_1$};
\node[ell] (o_2)at (-1.75,-1.5) {$\omega_2$};
\node[ell] (o_3)at (-0.25,-1.5) {$\omega_3$};
\node[ell] (o_4)at (1.25,-1.5) {$\omega_4$};
\node at (0.5,1.5) {$\Gamma_2$};
\draw [->] (A_) to (B_1);
\draw [->] (A_) to (B_2);
\draw [->] (B_1) to (o_1);
\draw [->] (B_1) to (o_2);
\draw [->] (B_2) to (o_3);
\draw [->] (B_2) to (o_4);
\end{tikzpicture}
\hspace{0.25cm}
\begin{tikzpicture}
\node[ell] (A__)at (-14.5,1.5){$A$};
\node[ell] (B__)at (-13.5,0) {$B$};
\node[ell] (o__1)at (-15.5,-1.5) {$\omega_1$};
\node[ell] (o__2)at (-12.75,-1.5) {$\omega_2$};
\node at (-13.5,1.5) {$\Gamma_3$};
\draw [->] (A__) to (B__);
\draw [->] (B__) to (o__1);
\draw [->] (B__) to (o__2);
\draw [->] (A__) to (o__1);
\end{tikzpicture}

\vspace{1cm}

\begin{subfigure}{0.3\linewidth}
\centering
$\begin{array}{|c|c|}
    \hline
    \omega_1 & \omega_1 \\ \hline
    \omega_2 & \omega_3 \\ \hline
\end{array}$
\caption*{$g_1$}
\end{subfigure}%
\hfill
\begin{subfigure}{0.2\linewidth}
\centering
$\begin{array}{|c|c|c|c|}
    \hline
    \omega_1 & \omega_1 & \omega_2 & \omega_2 \\ \hline
    \omega_3 & \omega_4 & \omega_3 & \omega_4 \\ \hline
\end{array}$
\caption*{$g_2$}
\end{subfigure}%
\hfill
\begin{subfigure}{0.3\linewidth}
\centering
$\begin{array}{|c|c|}
    \hline
    \omega_1 & \omega_1 \\ \hline
    \omega_1 & \omega_2 \\ \hline
\end{array}$
\caption*{$g_6$}
\end{subfigure}
\caption{
Acyclic deterministic graphical game structures and corresponding game forms}
\label{f2}
\end{figure}

\smallskip 

Acyclic DGGS  $\Gamma_1$  in Figure \ref{f2} generates game form  $g_1$.
Recall game $(g_1;u,w)$ from Section \ref{ss08a} with 
$u(\omega_2) > u(\omega_1) > u(\omega_3) \text{ and } w(\omega_2) > w(\omega_3).$
Note that the Backward Induction NE 
(see \cite{Gal53,Kuh53} and also \cite{Gur17}) is NE-A and  
is not Pareto-optimal. 
In general, this NE may differ from both, NE-A and NE-B. 

Acyclic DGGS  $\Gamma_3$  in Figure \ref{f2} generates game form  $g_6$;
see also Figure~\ref{f1}.

Clearly, in absence of dicycles in $\Gamma$,
the concepts of DGGS and MSDGGS coincide.
It is also clear that an acyclic DGGS is a special case of MSDGGS.
Thus, properties (I, II, III) required from an oracle hold for both.

\subsubsection*{Cyclic deterministic graphical game structures}
The outcomes of MSDGGS  are all its non-transient  SCCs.
In particular, each terminal position is an outcome.
Let us now assume that every simple dicycle is a separate outcome
(and each terminal remains an outcome as well).
Such  DGGSs, called {\em cyclic}, were studied in \cite{BGMS07};
some special cases were considered earlier \cite{GG91,GG91a,GG92}. 
Cyclic DGGS can also serve as oracles generating game forms;
see examples in \cite[Figures 1 and 2]{BGMS07};
compare examples 3 and 4 in  \cite[Figure 2]{BGMS07}  
with game forms  $g_4$  and $g_5$ in Figure 1.

Game forms generated by the cyclic DGGS
may be not tight; see Figure~1 in \cite{BGMS07}.
In other words, property (II) fails for the corresponding oracles, in general.
Yet, it holds in some important special cases.

A digraph  $G = (V,E)$  is called {\em symmetric}
if  $(v,v') \in E$  whenever  $(v',v) \in~E$.
Cyclic DGGS on symmetric digraphs are called  {\em symmetric}.
Symmetric Cyclic DGGSs satisfying (II) are called {\em solvable} 
and explicitly characterized in \cite[sections  1-3 and  Theorems 1-3]{BGMS07}. 
It follows from results of \cite{BGMS07} that
(III) also holds for solvable cyclic symmetric DGGS.
Hence, Theorem \ref{t2} is applicable.

\subsection{Jordan oracle; choosing Battlefields in Wonderland}
\label{ss41}
Wonderland is a subset of the plane homeomorphic to the closed disc.
Wlog, we can consider a square  $Q$ with the sides $N, E, S, W$. 
Let us partition $Q$  into areas
$\Omega = \{\omega_1, \dots, \omega_p\}$  each of which
is homeomorphic to the closed disc, too.
Every two distinct areas  $\omega_i, \omega_j \in \Omega$  are either disjoint or
intersect in a set homeomorphic to a closed interval that contains more than one point.
Equivalently, we can require that the borders
of the areas in $Q$  form a regular graph of degree $3$.
(Note that four vertices of the square are not vertices of this graph.)
Two examples are given in Figures \ref{f3} and \ref{f4}.

\begin{remark}
Consider game form $g_5$ in Figure \ref{f1} and
merge outcomes $\omega_5$ and $\omega_6$ in it getting $g_5'$.
(This operation respects tightness).
Note that $g_5' \in G$, where $G$ is the game correspondence given in Figure \ref{f3}.
See also \cite[Figure 4]{BGMS07}, 
where  $g_5$ also appears as the  normal form of a cyclic game form.
\end{remark}

\begin{figure}
\begin{center}
\begin{tikzpicture}
\draw (-1.5,-1.5) -- (-1.5,1.5) -- (1.5,1.5) -- (1.5,-1.5) -- (-1.5,-1.5);
\draw (1.5,0)--(0.5,0);\draw (-1.5,0)--(-0.5,0);\draw(0,1.5)--(0,0.5);\draw(0,-1.5)--(0,-0.5);
\draw(0.5,0)--(0,0.5)--(-0.5,0)--(0,-0.5)--(0.5,0);
\node at (0,0) {$\omega_5$};\node at (-1,-1) {$\omega_3$};\node at (1,-1) {$\omega_4$};\node at (-1,1) {$\omega_1$};\node at (1,1) {$\omega_2$};
\node at (2,0) {E};\node at (0,2) {N};\node at (-2,0) {W};\node at (0,-2) {S};
\end{tikzpicture}
\end{center}
\vskip 1cm
\begin{center}
\begin{tabular}{|c|c|c|c|}
\hline
$\omega_1$ & $\omega_2$ & $\omega_1$ & $\omega_2$
\\ \hline
$\omega_3$ & $\omega_4$ & $\omega_4$ & $\omega_3$
\\ \hline
$\omega_1$ & $\omega_4$ & $\omega_1 \omega_4 \omega_5$ & $\omega_5$
\\ \hline
$\omega_3$ & $\omega_2$ & $\omega_5$ & $\omega_2 \omega_3 \omega_5$ \\
\hline
\end{tabular}
\end{center}
\bigskip
\caption{
The Jordan game correspondence of the map of Wonderland.}
\label{f3}
\end{figure}

The following interpretation was suggested in \cite{GK18}.
Two players, Alice Tweedledee and Bob Tweedledum, agreed to have a battle.
The next thing to do is to agree on a battlefield, 
which should be an area $\omega \in \Omega$.
The strategies  $x \in X$  of Alice
are all (inclusion-minimal) subsets
$x \subseteq \Omega$ 
connecting  $W$  and  $E$,
Respectively, the strategies  $y \in Y$  of Bob
are all (inclusion-minimal) subsets $y \subseteq \Omega$
connecting  $N$  and  $S$.

\begin{proposition}
Any two such subsets $x$ and $y$ intersect. 
\end{proposition}

\proof 
It follows the Jordan curve theorem and the fact 
that all vertices in the square are of degree 3 
(except its four corners, which are of degree 2). 
Note that  $x$ and $y$  might be disjoint 
if we allow vertices of degree $4$ or more.  
\qed 

\medskip 

Intersection  $x \cap y$  may contain several areas of  $\Omega$.
Thus, a game correspondence
$G : X \times Y \rightarrow 2^\Omega \setminus \{\emptyset\}$ is defined.

\begin{proposition}
Game correspondence  $G$  is tight.
\end{proposition}

\proof 
Again, it follows from the Jordan curve theorem  and 
the assumption that all vertices in the square are of degree 3. Choose an arbitrary  $g \in G$ and consider a $\pm 1$ game $(g;\Omega_A,\Omega_B)$ 
determined by a partition $\Omega = \Omega_A \cup  \Omega_B$. 
Then, from the following two options exactly one holds:


\smallskip 
\noindent 
(a) areas from  $\Omega_A$  connect  W and E; \; 
(b)  areas from  $\Omega_B$  connect  N and S.  \qed

\medskip

The above observations imply that Jordan oracle  $\cO$
satisfies requirements  (I) and  (II). It remains to verify  (III),

\begin{proposition}
By using oracle  $\cO$, one can decide 
whether (a) or (b)  holds and 
find corresponding  $x$ or $y$, respectively, 
in time linear in $|\cO|$.  
\end{proposition}

\proof 
Consider all areas from  $\Omega_B$  boarding  N,
then add all areas from $\Omega_B$  boarding these areas, etc.
Such iterations will stop in time linear in  $|\cO|$ 
either reaching  S (then, obviously, (b) holds) or not
(then (a) holds, again by the Jordan curve theorem).
Moreover, in the first case we obtain
a set of areas  $y'$  from  $\Omega_B$  connecting  N and S;
in the second case - a set of areas
$x'$  from  $\Omega_A$  connecting  W and  E.
The former strategy  $y'$  is obtained explicitly;
the latter one,  $x'$, is easy to  construct.
To do so, denote by  $\Omega'_B$  the set of areas
obtained in the course of iterations.
It does not  reach  S.
Hence, the areas from  $\Omega_A$  that border  $\Omega'_B$
connect  W and E, by the Jordan curve theorem once more. \qed

\smallskip 

This case is realized in Figure \ref{f4}; Alice wins.

\smallskip 

\begin{figure}
\centering
\begin{tikzpicture}
\draw[ultra thick](0,0) rectangle (7,-7);

\node at (7.75,-3.5) {E};\node at (3.5,0.5) {N};\node at (-0.75,-3.5) {W};\node at (3.5,-7.5) {S};

\filldraw[ultra thick,fill=lightgray] (1.75,-0.5)--(2.25,-0.5)--(2.5,-1)--(2.25,-1.5)--(1.75,-1.5)--(1.5,-1)--cycle;

\filldraw[ultra thick,fill=lightgray] (4.5,-4.5)--(5.5,-4.5)--(6,-5.5)--(5.5,-6.5)--(4.5,-6.5)--(4,-5.5)--cycle;

 \filldraw[ultra thick,fill=lightgray] (3,0) -- (3.5,-0.5) -- (4.5,-0.5) -- (5,0)--cycle;
\filldraw[ultra thick,fill=lightgray] (0,-1) -- (0,-3) -- (0.25,-3.25) -- (0.5,-3) -- (0.5,-2)--cycle;
\draw[ultra thick] (0.5,-3) -- (0.5,-2)-- (1,-3)--cycle;
\draw[ultra thick] (0,-3) -- (0,-4)-- (0.5, -4.5)--(1,-4)--cycle;
\filldraw[ultra thick,fill=lightgray] (0.25,-3.25) -- (1,-4)-- (1.25,-3.75)--(1,-3)--(0.5,-3)--cycle;
\filldraw[ultra thick,fill=lightgray] (0.5,-4.5) -- (1.25,-3.75)-- (1.5,-4.5)--(1,-5)--cycle;
\filldraw[ultra thick,fill=lightgray] (1.5,-4.5)--(2,-5)--(2,-6)--(1,-5)--cycle;
\draw[ultra thick] (2,-5)--(2,-5.5)--(2.75,-4.5)--(2.5,-4)--cycle;
\filldraw[ultra thick,fill=lightgray] (2,-5.5)--(2.75,-4.5)--(3,-5)--(2,-6)--cycle;
\filldraw[ultra thick,fill=lightgray] (2.5,-4)--(3,-5)--(3.5,-4.5)--(3,-3.75)--(2.75,-3.5)--cycle;
\draw[ultra thick] (2.75,-3.5)--(3,-3)--(3.5,-3.5)--(3,-3.75)--cycle;
\filldraw[ultra thick,fill=lightgray] (3.5,-3.5)--(3,-3.75)--(3.5,-4.5)--(4,-4)--cycle;
\filldraw[ultra thick,fill=lightgray] (3,-3)--(3.5,-3.5)--(3.75,-3.75)--(4.25,-3)--(4,-3)--(3.25,-2.5)--cycle;
\draw[ultra thick] (4.25,-3)--(3.75,-3.75)--(4,-4)--(4.5,-3)--cycle;
\filldraw[ultra thick,fill=lightgray] (3.25,-2.5)--(4,-2.5)--(4,-3)--cycle;
\filldraw[ultra thick,fill=lightgray] (3,-2)--(4,-2)--(4,-2.5)--(3.25,-2.5)--(3.1,-2.8)--cycle;
\draw[ultra thick] (4,-2)--(4.5,-2)--(4.5,-3)--(4,-3)--cycle;
\filldraw[ultra thick,fill=lightgray] (4.5,-3)--(4.5,-2.5)--(5,-3.5)--(5,-4.25)--(4.25,-3.5)--cycle;
\draw[ultra thick] (4.5,-3.75)--(4.25,-3.5)--(4.125,-3.75)--cycle;
\filldraw[ultra thick,fill=lightgray] (4.5,-3.75)--(4.125,-3.75)--(4,-4)--(3.75,-4.25)--(5,-4.25)--cycle;
\draw[ultra thick] (5,-3.5)--(5.5,-2)--(6,-3.5)--(5,-4)--cycle;
\draw[ultra thick] (19/4,-3)--(31/6,-3)--(5,-3.5)--cycle;
\filldraw[ultra thick,fill=lightgray] (19/4,-3)--(31/6,-3)--(5.5,-2)--(5,-1)--(4.5,-2)--(4.5,-2.5)--cycle;
\filldraw[ultra thick,fill=lightgray] (5.25,-1.5)--(6,-1.5)--(6,-3)--(35/6,-3)--(5.5,-2)--cycle;
\draw[ultra thick] (5,-1)--(5.5,-0.5)--(6,-1)--(6,-1.5)--(5.25,-1.5)--cycle;
\draw[ultra thick] (6,-3)--(35/6,-3)--(6,-3.5)--(7,-3)--(7,-2)--cycle;
\filldraw[ultra thick,fill=lightgray] (6,-3)--(6,-2)--(6.5,-1.5)--(6.5,-2.5)--cycle;

\filldraw[ultra thick,fill=lightgray] (6.5,-7/4)--(7,-1.5)--(7,-1)--(6.5,-1)--(6.5,0)--(6,0)--(5.5,-0.5)--(6,-1)--cycle;
\filldraw[ultra thick,fill=lightgray] (6.5,0)--(6.75,0)--(6.75,-0.5)--(6.5,-1)--cycle;
\filldraw[ultra thick,fill=lightgray] (6.75,0)--(6.75,-0.5)--(7,-0.5)--(7,0)--cycle;

\filldraw[ultra thick] (4,-5.5)--(1,-7);
\filldraw[ultra thick] (4.5,-6.5)--(5,-7);
\filldraw[ultra thick] (5.5,-6.5)--(5.5,-7);
\filldraw[ultra thick] (6,-5.5)--(7,-6);
\filldraw[ultra thick] (4.5,-4.5)--(3.25,-4.75);
\filldraw[ultra thick] (5.5,-4.5)--(7,-5);
\filldraw[ultra thick] (0,-5)--(0.5,-7);
\filldraw[ultra thick] (6.5,-3.25)--(6.5,-29/6);
\filldraw[ultra thick,fill=lightgray] (0.25,-6)--(1,-6)--(1/3,-19/3)--cycle;
\filldraw[ultra thick,fill=lightgray] (1,-7)--(3,-6)--(4,-7)--cycle;
\draw[ultra thick] (1,-6)--(2,-6.5);

\draw[ultra thick] (2.5,-1)--(3.5,-0.5);
\draw[ultra thick] (2.25,-1.5)--(2,-2);
\draw[ultra thick] (0.25,-1.5)--(2,-2);
\draw[ultra thick] (1.75,-1.5)--(1,-1);
\draw[ultra thick]
(1,-1)--(1.5,-1);
\draw[ultra thick]
(1,-1)--(1,0);
\draw[ultra thick]
(2,-2)--(4.5,-0.5);
\draw[ultra thick]
(1.75,-0.5)--(2,-0.25);
\draw[ultra thick]
(2,-0.25)--(2.25,-0.5);
\draw[ultra thick]
(2,-0.25)--(2,0);
\draw[ultra thick]
(3,-2)--(0.75,-2.5);
\draw[ultra thick]
(3.5,-2)--(5.5,0);
\draw[ultra thick](43/10,-12/10)--(4.75,-1.5);
\filldraw[ultra thick,fill=white](6.75,-1)--(7,-1)--(7,-1.5)--(6.75,-3.25/2)--cycle;
\end{tikzpicture}
    \caption{Gray and white areas are in $\Omega_A$ and $\Omega_B$, respectively. Alice wins.}
    \label{f4}
\end{figure}
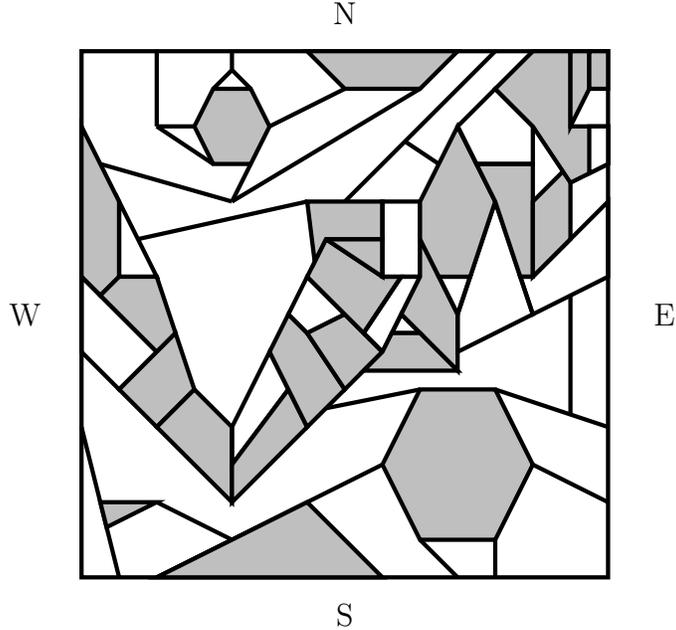

\begin{remark}
It is not necessary to restrict ourselves by minimal strategies. 
In linear time we can reduce arbitrary strategy 
(set)  $x'$  of Alice  to an inclusion-minimal set  $x$
connecting  W  and  E,   
thus, getting minimal strategies of Alice. 
To do so, we eliminate areas from  $x'$ 
one by one until  (a)  still holds.
We require inclusion-minimality
of subsets  $x \in \Omega$  just to reduce the number of strategies
(which may still remain exponential  in $|\cO|$). 
Of course, the same is true for Bob's strategies. 
\end{remark}

\subsection{Monotone bargaining schemes}
\label{ss43}
The following oracle was introduced in \cite{GK18}.
Two players, Alice and Bob, possess items
$A = \{a_1, \dots, a_m\}$  and  $B = \{b_1, \dots, b_n\}$, respectively.
Both sets are ordered:
$a_1 \prec \dots \prec a_m$  and  $b_1 \prec \dots \prec b_n$.
Both players know both orders.

The direct product
$\Omega = A \times B = \{(a, b) \mid  a \in A, b \in B\}$
is the set of {\em outcomes}.

Alice's strategies are monotone non-decreasing mappings  $x: A \rightarrow B$
(that is, $x(a) \geq  x(a')$  whenever  $a > a'$)
showing that she is ready to exchange  $a$  for  $x(a)$   for any  $a \in A$.
Similarly, Bob's strategies are monotone non-decreasing mappings  $y: B \rightarrow A$
(that is, $y(b) \geq  y(b')$  whenever  $b > b'$)
showing that he is ready to exchange  $b$  for  $y(b)$  for any  $b \in B$.

It is not difficult to compute 
the numbers of strategies and outcomes:
\begin{equation}
\label{eq1}
|X| = \binom {m+n-1}{m}, \; |Y| = \binom {m+n-1}{n}; \;\; |\Omega| = |A \times B| = mn.
\end{equation}

Given a situation $(x,y)$,
an outcome  $(a,b) \in \Omega$  is called  a {\em deal}
(in this situation) if  $x(a) = b$  and  $y(b) = a$.
Denote by  $G(x,y) \subseteq \Omega$  the set of all deals in the situation $(x,y)$.
We will show that $G(x,y) \neq \emptyset$. 
Yet, $G(x,y)$  may contain several deals.

This construction is called a {\em monotone bargaining (MB) scheme}.
It can be viewed as an oracle $\cO$
generating game correspondence  $G : X \times Y \rightarrow 2^\Omega \setminus \{\emptyset\}$.
By \eqref{eq1}, requirement (I) holds for  $\cO$.

Note that  $G = G_{m,n}$  is uniquely defined by  $m$  and  $n$.
A game form  $g \in G$  is called  an {\em MB game form}.
For example, if  $m=n=3$
then  $|X| = |Y| = 3$  and we obtain game form  $g_4$  in Figure 1; 
game correspondence $G(x,y)$  is given in \cite[Figure 1a]{GK18}.  

The following interpretation was suggested in \cite{GK18}.
Alice and Bob are dealers possessing the sets of objects
$A$  and  $B$, respectively, and
a deal  $(a,b) \in A \times B$  means that
they exchange  $a$  and  $b$.
They may be art-dealers, car dealers; or one
of them may be just a buyer with a discrete budget.
For example, $A = \{a_1, \dots, a_m\}$  and  $B = \{b_1, \dots, b_n\}$
may be paintings or sculptures
ordered in accordance with their age (not price or value).

\medskip

To any pair of mappings
$x : A \rightarrow B$  and  $y : B \rightarrow A$
(not necessarily  monotone non-decreasing)
let us assign a bipartite digraph  $\Gamma = \Gamma(x,y)$
on the vertex-set  $A \cup B$  as follows:  $[a,b)$
(respectively, $[b,a)$) is an arc of $\Gamma(x,y)$  whenever  $x(a) = b$
(respectively, $y(b) = a$).

Some visualization helps.
Embed $\Gamma(x,y)$ into a plane;
putting ordered  $A$ and  $B$  in two parallel columns.
Two arcs corresponding to  $x$  may have a common head, but not tail.
Furthermore, they cannot cross if mapping $x$  
is monotone non-decreasing. Similarly  for  $y$.
By construction, digraph  $\Gamma$  is bipartite, with parts  $A$  and $B$.
Hence, every dicycle in  $\Gamma$  is even.
There is an obvious one-to-one correspondence between
the dicycles of length $2$  
in $\Gamma(x,y)$ and the deals of  $G(x,y)$.

\begin{proposition}
\label{p-deals}
For each situation $(x,y)$ its  digraph  $\Gamma(x,y)$
contains at least one dicycle of length  $2$
(a deal) and cannot contain longer dicycles.
\end{proposition}

\proof
For any initial vertex  $v \in A \cup B$,
strategies  $x$  and  $y$  uniquely
define an infinite walk from  $v$,
which is called a play.
Since sets $A$  and  $B$ are finite and there are no terminals, this play is a {\em lasso}:
it consists of an initial directed path, which may be empty, and a dicycle  $C$  repeated infinitely.
Furthermore, $C$  must be a dicycle of length $2$ 
whenever mappings  $x$  and  $y$  are monotone non-decreasing.
Indeed, if $C$ is longer than $2$ then 
crossing arcs appear and, hence, 
either  $x$, or $y$, or both are not monotone,
\qed

\medskip

Consider a $\pm 1$ MB game  $(g;\Omega_A,\Omega_B)$,
where  $g = g(\cO)$  is an MB game form generated by an MB scheme $\cO$.
As we already mentioned, requirement  (I)  holds for  $\cO$.
The following statement shows that  (II)  and  (III) hold as well.

\begin{proposition}
\label{pMBS}
Game form  $g = g(\cO)$  is tight and each $\pm 1$ MB game
$(g;\Omega_A,\Omega_B)$  can be solved in time polynomial in $|\cO| = mn$.
\end{proposition}

The first part was already proven in \cite{GK18}.
Yet, here we provide a much shorter proof.

\proof

For the sake of simplicity, we will slightly abuse notation 
writing that both directed edges $[a,b)$  and  $[b,a)$  are
in  $\Omega_A$ or in $\Omega_B$  
whenever the corresponding deal $(a,b)$ is in $\Omega_A$ or in $\Omega_B$, respectively.

Consider complete bipartite symmetric digraph  $\Gamma$  on  $m+n$ vertices
$A = \{a_1, \dots, a_m\}, B = \{b_1, \dots, b_n\}$,
and with $2mn$ directed edges
$\{[a_i,b_j), [b_j,a_i) \mid i = 1, \dots, m; j = 1, \dots, n\}$. 
The following two  statements are obvious: 

\smallskip 

(a) Alice wins if she has a monotone non-decreasing strategy
$x^* : A \rightarrow B$  such that $[a, x^*(a)) \in \Omega_A$  for all $a \in A$.

\smallskip 

(b) Bob wins if  he has a monotone non-decreasing strategy
$y^* : B \rightarrow A$  such that $[b,y^*(b)) \in \Omega_B$  for all $b \in B$.

\smallskip 

Indeed, it is easily seen that  $x^*$  and $y^*$  are
the winning strategies of Alice and Bob, respectively.
It is enough to show that
$g(x^*,y) \in \Omega_A$  for any $y \in Y$.
Recall the proof of Proposition  \ref{p-deals}:
Fix  $x^*$, choose an arbitrary $y \in Y$, and consider the play
$P = P(x^*,y)$  beginning from an arbitrary initial position $v \in A \cup B$.
By Proposition  \ref{p-deals}, $P$  is a lasso resulting in a $2$-cycle  $(a,b)$.

The corresponding deal $(a,b) \in \Omega_A$, in case (a),
for any  $y$,  by the choice of  $x^*$, and hence, Alice wins.
Similarly, $g(x,y^*) \in \Omega_B$  in case (b)  for any $x \in X$,
by the choice of  $y^*$, and Bob wins.

Obviously, (a) and (b)  cannot hold simultaneously,
since otherwise $(a,b) \in  \Omega_A \cap \Omega_B$, which is a contradiction,
since  $\Omega = \Omega_A \cup \Omega_B$  is a partition.

\medskip

Let us show that either (a) or (b) holds 
(in other words,  $g$  is tight, which implies (II)).
The  proof will be constructive:
we obtain either $x^*$ satisfying (a) or $y^*$ satisfying (b)
in time polynomial in $mn$  (which in its turn, implies (III)). 

We will construct a play  $P$  by the following greedy iterative algorithm.
Let  $a^1 = a_1$  be an initial position of  $P$.
(We use superscripts to number iterations.)

If $[a^1,b) \in \Omega_B$  for all  $b \in B$  then Bob wins.
(His winning strategy  $y^*$ is defined by:
$y^*(b) = a^1$  for all $b \in B$.
Then $[y^*(b),b) \in \Omega_B$ for all $b \in B$ and (b) holds.)
Otherwise, denote by  $b^1$ the 
(unique) minimal $b \in B$ such that $[a^1,b) \in \Omega_A$.
Then, by definition, $[b^1, a^1) \in \Omega_A$  too. 
Furthermore, by this choice of  $b^1$, we have:
$[b, a^1) \in \Omega_B$  for all  $b \prec b^1$,
while $[b^1, a^1) \in \Omega_A$.

If $[b^1,a) \in \Omega_A$  for all $a \succeq a^1$  then Alice wins.
(Her winning strategy  $x^*$ is defined by:
$x^*(a) = b^1$  for all $a \in A$.
Then $[a,x^*(a)) \in \Omega_A$ for all $a \in A$.)
Otherwise, denote by  $a^2$ the (unique) minimal $a \in A$
such that $[b^1,a) \in \Omega_B$.
Then, by definition, $[a^2, b^1) \in \Omega_B$  too. 
Furthermore, by the choice of  $a^2$, we have:
$[a, b^1) \in \Omega_A$  for all $a \prec a^2$,
while $[a^2, b^1) \in \Omega_B$.

The general $k$-th step of this greedy recursion is as follows.

If  $[a^k,b) \in \Omega_B$  for all $b \succeq b^{k-1}$  then Bob wins.
(His  winning strategy   $y^*$  is defined by:
$y^*(b) = a^i$  for each  $b$  such that $b^i \succ b \succeq b^{i-1}$,
for  $i   = 1, \dots, k$, assuming conventionally that
$b \succ b^0$  holds for all $b \in B$).

Otherwise, denote by  $b^k$ the (unique) minimal $b \in B$
such that $b \succ b^{k-1}$ and  $[a^k,b) \in \Omega_A$.
Then $[b^k,a^k) \in \Omega_A$  too.

Furthermore, by the choice of  $b^k$, we have:
$[b, a^k) \in \Omega_B$  for all $b$  such that
$b^k  \succ b \succeq b^{k-1}$, while $[b^k, a^k) \in \Omega_A$.

If $[b^k,a) \in \Omega_A$  for all $a \succeq a^k$  then Alice wins.
(Her winning strategy  $x^*$ is defined by:
$x^*(a) = b^j$  for each  $a$  such that $a^{j+1} \succ a \succeq a^j$,
for  $j = 1, \dots, k$, assuming conventionally that
$a^{k+1} \succ a$  holds for all $a \in A$.)

Otherwise, denote by  $a^{k+1}$ the 
(unique) minimal $a \in A$ such that 
$[b^k,a) \in \Omega_B$. Then $[a^{k+1}, b^k) \in \Omega_B$, too.

Furthermore, by the choice of  $a^{k+1}$, we have:
$[a, b^k) \in \Omega_A$  for all  $a$  such that
$a^{k+1} \succ a \succeq a^k$, while $[a^{k+1}, b^k) \in \Omega_B$.

\medskip
After each iteration $a^k$ (respectively, $b^k$) both Alice and Bob 
have winning moves in all positions $a \prec a^k$ and $b \preceq b^{k-1}$ 
(respectively,  $a \preceq a^k$ and $b \prec b^k$). Since sets  $A$ and $B$ are finite,
the procedure will stop on some iteration either 
$a^{k^*} \prec a_m$ or $b^{k^*} \prec b_n$,
indicating that Bob or, respectively, Alice wins.

Furthermore, we obtain his or her winning strategy in time linear in $mn$.

\medskip

The following slightly different procedure can be applied too.
First, we start looking for a winning strategy  $x^*$  for Alice.
Consider successively  $a_1, a_2, \dots$  and construct
(again recursively and greedily) her monotone non-decreasing
strategy  $x^*$  as follows:
$x^*(a_i) = b^i$  such that $[a_i,b^i) \in \Omega_A$, $b^i \succeq  b^{i-1}$, and
$b^i$  is the minimal element of  $B$  satisfying these two properties.
If this will work for all  $i = 1, \dots, m$  then Alice wins and
we  obtain her winning strategy  $x^*$  satisfying (a).
Otherwise, if the procedure stops on some $i < m$
(no required  $b^i$  exists for  $a_i$) then Bob wins.
His winning strategy  $y^*$  satisfying  (b) is defined as follows:
$y^*(b) =  a^i$  for all  $b$  such that $b^{i-1} \preceq b \prec b^i$,
where  $a_i$ is the smallest  $a$  such that  $x^*(a) = b^i$,
for $i = 1,2, \dots$ 
By convention, $b^0 \prec b$  for all $b \in B$. \qed

\medskip

Thus, requirements (I,II,III) hold for the MB schemes and, 
hence, Theorem \ref{t2} is applicable.

\subsection{Veto voting schemes}
\label{ss44}
Two voters (players), Alice and Bob
choose among candidates (options, outcomes) $\Omega = \{\omega_1, \dots, \omega_p\}$.
They are assigned some positive integer veto powers
and given  $\mu_A$  and  $\mu_B$   veto cards, respectively.
Each candidate  $\omega \in \Omega$  is assigned 
an integer positive veto resistance  $\lambda_\omega$.
We assume that

\begin{equation}
\label{+1}
\mu_A + \mu_B + 1 = \lambda_{\omega_1} + \dots + \lambda_{\omega_p}.
\end{equation}

A strategy of a voter is an arbitrary distribution
of her/his veto cards among the candidates.
Given a pair of strategies  $x$  and $y$, a candidate
$\omega \in \Omega$  who got at least  $\lambda_\omega$  veto cards 
(from Alice and Bob together) is vetoed.
From the set  $G(x,y)$  of all not vetoed candidates one $g(x,y) \in G(x,y)$  is elected.
By \eqref{+1}, $G(x,y) \neq \emptyset$.
Thus, we obtain a {\em veto voting} (VV) scheme  $\cO$,
VV game form  $g =  g(\cO)$, and VV game correspondence  $G = G(\cO)$;  
see, for example, \cite{Gur08},\cite[Chapter 6]{Mou83},\cite[Chapter 5]{Pel84} for more details.

By construction, VV schemes are oracles satisfying (I).
For example, game form   $g_3$  in Figure 1 corresponds to the VV scheme defined by
$$\mu_A = \mu_B = \lambda_{\omega_1} = \lambda_{\omega_2} = \lambda_{\omega_3} = 1.$$

\medskip

Let us show  that requirements (II) and (III) also hold for VV schemes.

\begin{proposition}
Each game form  $g$ defined by a VV scheme satisfying (\ref{+1}) is tight.
Furthermore, every $\pm 1$ game  $(g;\Omega_A,\Omega_B)$  can be solved in time linear in
$|\cO| = \log (\mu_A \mu_B \prod_{\omega \in \Omega} \lambda_\omega)$.
\end{proposition}

\proof
To  see  this, consider a $\pm 1$ game  $(g;\Omega_A,\Omega_B)$.
By  (\ref{+1}), from two options,
(a) Alice can veto  $\Omega_B$  and
(b) Bob can veto  $\Omega_A$, exactly one holds.
Alice or Bob wins in case of (a) or (b), respectively.
Given numbers  $\mu_A, \mu_B$,  and  $\lambda_\omega, \; \omega \in \Omega$,
one can decide in linear time whether (a) or  (b)  holds.
In each case the winning strategy of Alice or Bob is straightforward:
just veto all opponent's candidates, $\Omega_B$ or $\Omega_A$, respectively.
\qed

Thus, the VV oracles satisfy (I,II,III)  and Theorem \ref{t2} is applicable.

\subsection{Tight game correspondences and forms of arbitrary monotone properties}
\label{ss25}  
The most general setting is defined as follows.
Given a finite ground set $\Omega$, 
consider a family of its subsets $\cP \subseteq 2^\Omega$.
Standardly, we call $\cP$ a {\em property} and say that a subset  
$\Omega' \subseteq \Omega$   satisfies  $\cP$  or not 
if $\Omega' \in \cP$  or  $\Omega' \not\in \cP$, respectively.    
Property $\cP$  is called {\em  inclusion monotone non-decreasing}  
(or simply  {\em monotone}, for short) if
$\Omega'' \in \cP$  implies  $\Omega' \in \cP$  
whenever  $\Omega'' \subseteq \Omega' \subseteq \Omega$.
We restrict ourselves to monotone properties.  

Define the sets of strategies  $X$  of Alice and  $Y$ of Bob as follows:

\medskip 

$x \in X$   is any (inclusion minimal) subset  
$\Omega_A \subseteq  \Omega$  such that  $\Omega_A \in \cP$; 

\smallskip 

$y  \in Y$   is any  (inclusion minimal) subset 
$\Omega_B \subseteq \Omega$ such that $\Omega \setminus \Omega_B \not\in \cP$.

\medskip 

The restriction in parenthesis does not matter, it can be waved or kept.
In the latter case, sets  $X$  and  $Y$  are significantly reduced.  

Define a game correspondence  $G = G(\cP)$  by setting  $G(x,y) = x \cap y$.
It is  both obvious and well-known that   
$G(x,y) \neq \emptyset$  for any  $x  \in X, y  \in Y$ and,    
moreover, $G$  is tight.
Hence, any game form  $g \in G$  is tight too.

Thus, (I) and (II)  hold automatically
whenever a monotone property  $\cP$  is  given by an oracle  $\cO(\cP)$.
Yet, (III) must be required in addition.
In other words, $\cO(\cP)$  must be a polynomial membership oracle, which 
for a  given subset  $\Omega' \subseteq \Omega$, decides if  $\Omega' \in \cP$  
in time polynomial in $|\Omega| + |\cO(\cP)|$.  

It is easily seen that this general setting includes in particular
all four examples of oracles given in this section before; 
see more examples in \cite{BGEK02,Gur18a}.

\section*{Acknowledgement}
This research was prepared within the framework 
of the HSE University Basic Research Program.  
The authors are thankful to Endre Boros and 
to three anonymous reviewers for helpful remarks and suggestions.

\end{document}